\documentclass[twocolumn,amsmath,amssymb,nofootinbib,prd,aps,10pt,floatfix]{revtex4-2}
\usepackage[utf8]{inputenc}
\usepackage{graphicx, float,amsmath, amsmath, amssymb}

\usepackage{amsfonts}
\usepackage{bm}
\usepackage{dcolumn}
\usepackage[braket]{qcircuit}
\usepackage[export]{adjustbox}
\usepackage{multirow}
\usepackage{color}
\usepackage{caption}
\usepackage{subcaption}
\usepackage{tensor}
\usepackage{listings}
\usepackage{xcolor}

\definecolor{codegreen}{rgb}{0,0.6,0}
\definecolor{codegray}{rgb}{0.5,0.5,0.5}
\definecolor{codepurple}{rgb}{0.58,0,0.82}
\definecolor{backcolour}{rgb}{0.95,0.95,0.92}

\lstdefinestyle{mystyle}{
    backgroundcolor=\color{backcolour},   
    commentstyle=\color{codegreen},
    keywordstyle=\color{magenta},
    numberstyle=\tiny\color{codegray},
    stringstyle=\color{codepurple},
    basicstyle=\ttfamily\footnotesize,
    breakatwhitespace=false,         
    breaklines=true,                 
    captionpos=b,                    
    keepspaces=true,                 
    numbers=left,                    
    numbersep=5pt,                  
    showspaces=false,                
    showstringspaces=false,
    showtabs=false,                  
    tabsize=2
}

\usepackage{hyperref}

\usepackage{tikz}
\usetikzlibrary{decorations.markings}
\usetikzlibrary{calc}

\tikzset{
    ingoing/.style={decoration={markings, mark=at position 0.5 with {\arrow{<}}}, postaction=decorate},
    outgoing/.style={decoration={markings, mark=at position 0.5 with {\arrow{>}}}, postaction=decorate}
}

\begin{document}

\title{From \texorpdfstring{$SU(2)$}{SU(2)} holonomies to holographic duality via tensor networks}
\author{Grzegorz Czelusta$^{1,2}$}
\author{Jakub Mielczarek$^{1}$}
\email{jakub.mielczarek@uj.edu.pl}
\affiliation{
$^{1}$Institute of Theoretical Physics, Jagiellonian University, 
{\L}ojasiewicza 11, 30-348 Cracow, Poland\\
$^{2}$Doctoral School of Exact and Natural Sciences, Jagiellonian University, 
{\L}ojasiewicza 11, 30-348 Cracow, Poland}
\date{\today}

\begin{abstract}
Tensor networks provide a powerful tool for studying many-body 
quantum systems, particularly making quantum simulations more 
efficient. In this article, we construct a tensor network 
representation of the spin network states, which correspond to 
$SU(2)$ gauge-invariant discrete field theories. Importantly, 
the spin network states play a central role in the Loop Quantum 
Gravity (LQG) approach to the Planck scale physics. Our focus is 
on the Ising-type spin networks, which provide a basic model 
of quantum space in LQG.  It is shown that the tensor network 
approach improves the previously introduced methods of constructing
quantum circuits for the Ising spin networks by reducing 
the number of qubits involved. It is also shown that the tensor 
network approach is convenient for implementing holographic 
states involving the bulk-boundary conjecture, which contributes 
to establishing a link between LQG and holographic duality. 
\end{abstract}

\maketitle

\section{Introduction}

Tensor networks have played an important role in recent years, 
not only in the quantum many-body systems but also in the 
context of gravity \cite{Orus:2013kga,Biamonte:2019uzx}. 
The relation to gravity is rooted in the hyperbolic geometry 
of the tensor networks, which correspond to certain discrete 
quantum systems. This has been studied especially in the context 
of the  Multi-scale Entanglement Renormalization Ansatz (MERA) 
tensor network \cite{Vidal:2007hda,Vidal:2008zz}, which is 
considered a discrete approximation to the anti-de Sitter 
(AdS) spacetime \cite{Swingle:2009bg}, in the context of 
Gauge/Gravity duality. 

On the other hand, different network structures, the
so-called spin networks, have been broadly investigated 
within the Loop Quantum Gravity (LQG) \cite{Rovelli:1997yv, Ashtekar:2004eh} approach to quantum gravity. Importantly, 
the spin networks describe the $SU(2)$-invariant states. 

In recent years, some research activity has been 
directed towards an attempt to develop quantum computing 
methods allowing for the future quantum simulations 
of the spin network states \cite{li2019quantum,Mielczarek:2018jsh,zhang2020observation,Mielczarek:2021xik,
Mielczarek:2021xik,Czelusta:2020ryq}. 
The analysis has been primarily focused on the case of 
4-valent spin networks with links described by holonomies 
belonging to the fundamental ($j=1/2$) representation of 
the $SU(2)$ group. Because in this case the $SU(2)$-invariant
subspace at the node is two-dimensional, 
$\text{dim}({\rm Inv}_{SU(2)}(V^{\otimes 4}_{1/2}))=2$, 
we refer to  \emph{Ising spin networks} in this case \cite{Feller:2015yta}. 

The basic quantum circuits for the nodes of Ising spin 
networks have been introduced in Refs. \cite{Mielczarek:2018jsh,
Mielczarek:2021xik,Czelusta:2020ryq}. 
In Ref. \cite{czelusta2023quantum}, a systematic framework enabling 
the construction of quantum circuits for the Ising spin networks 
has been introduced. The primary purpose of the current article 
is to develop and further improve this method using the tensor 
network methods. The current limited quantum computing resources 
justify such an attempt. Therefore, any reduction of quantum 
computation resources needed to simulate a given spin network 
state provides advancement toward large-scale quantum simulations 
of the states of LQG. Indeed, by explicitly constructing the tensor 
network representation of the Ising spin networks, we show that the 
approach provides a computational advantage. 

There is, however, also a second motivation for this work. 
Namely, as mentioned earlier, the tensor networks provide 
foundations to the holographic correspondence between the 
AdS spacetime and Conformal Field Theories (CFTs) know 
as AdS/CFT conjecture \cite{Maldacena:1997re}. This is 
because the structure of quantum entanglement, encoded via 
the tensor networks, attains geometric interpretation in 
the sense of a graph. The graph, in the continuous limit, 
corresponding, e.g., to conformal field theories, may 
converge to hyperbolic spacetimes such as the AdS. 

From this perspective, it is justified to explore the 
relation between the description of the bulk in LQG, where 
the spin networks are used, and the tensor network approach 
emerging in the context of the Gauge/Gravity duality, such as 
the AdS/CFT conjecture. Some first attempts to clarify the 
nature of this relation have already been made. In particular, 
in Ref. \cite{PhysRevD.95.024011}, it has been justified that 
the tensor networks can be considered coarse-grained 
spin networks in LQG. It has also been noticed that the 
Ising spin networks correspond to the Projected Entangled Pair 
States (PEPS) type tensor network \cite{Cirac:2020obd}, 
which exhibits the area law of entanglement entropy 
\cite{Czelusta:2020ryq}. Further developments concern the 
relation between the spin networks and the XZ-calculus, which, 
in the diagrammatic representation, leads to tensor networks 
\cite{east2023all}. 

The article is organized as follows: In Sec. \ref{Sec.Holonomies}
$SU(2)$ holonomies (Wilson lines), which are basic ingredients
of the spin networks, are introduced. Notably, an isomorphism 
between the holonomies considered as unitary maps and maximally 
entangled states is discussed. Specifically, the isomorphism 
allows for the raising and lowering of the tensor indices in the 
tensor network representation. Then, in Sec. \ref{Sec.BasicDictionary}
a basic dictionary relating spin networks and tensor networks 
is introduced in the case of the Ising spin networks. In Sec. 
\ref{Sec.QuantumCircuits} quantum circuit representation of the 
tensors discussed in the previous section is constructed. The 
quantum circuit representation allows for the utilization of the 
tensor network method to construct quantum circuits for arbitrary 
Ising spin networks. Explicit expressions for the five most essential  
quantum circuits considered in this section can be found in Appendix A. 
The holographic properties of the tensor 
network representation of the spin networks are explored in Sec. 
\ref{Sec.BulkBoundary}. The methods introduced in Sec. \ref{Sec.QuantumCircuits}
demand application of the projection operator, implying frequent 
repetition of a given quantum circuit. Some prospects for mitigating 
the problem by reducing the number of shots is discussed in 
Sec. \ref{Sec.ProjectionProblem}. In Sec. \ref{Sec.Applications}, 
we provide examples of the application of the introduced methods
to 5 nodes (pentagram) spin networks and bulk-boundary maps. 
Furthermore, the tensor network representation is used to analyze 
entanglement entropy for an exemplary bulk-boundary map. The results 
of our studies are summarised in Sec. \ref{Sec.Summary}, where some 
prospects for further research are also given. 

\section{Two shades of holonomies}
\label{Sec.Holonomies}

The links of the spin networks are associated with $SU(2)$ 
holonomies, which are unitary maps between the source and 
target Hilbert spaces. For general, irreducible representation 
$j$ of $SU(2)$, the holonomies are $(2j+1)\times (2j+1)$ 
matrices acting on spin-$j$ Hilbert spaces:
\begin{equation}
V_j = {\rm span}(\ket{j,-j},...,\ket{j,j}),   
\end{equation}
where $\ket{j,m}$ are eigenstates of the operator $\hat{J}_z$ 
operator, $\hat{J}_z\ket{j,m} = m \ket{j,m}$. The dimension 
of the Hilbert space is $d_j= {\rm dim} V_j = 2j+1$. 

Considering the source space $V_j^*$ and the target space 
being $V_j$, the holonomy operator can be expressed 
as:
\begin{equation}
\hat{h} = \sum_{m,m'} {D^m}_{m'}(h) \ket{j,m}_t\bra{j,m'}_s 
\in V_j \otimes V_j^*,
\label{HolonomyDef}
\end{equation}
where $ {D^m}_{m'}(h)$ are elements of the Wigner matrix 
corresponding to the holonomy $h \in SU(2)$. Therefore, 
the right-handed action of holonomy given by (\ref{HolonomyDef})
provides a mapping between the kets in the source and the 
target spaces: $\hat{h}: V_j \rightarrow V_j$.

It is instructive to consider the transformation of the 
holonomy operator under the change of bases at the 
source and the target spaces. The change of the basis state 
can be associated with the unitary operators $\hat{U}_s$ and 
$\hat{U}_t$, so that $\ket{j,m}' = \hat{U}_{s/t}\ket{j,m}$ 
and employing the components of the unitary matrix 
$\ket{j,m'}' = \sum_m U_{s/t,mm'}\ket{j,m}$.

Considering the holonomy after the change of basis, we find: 
\begin{align}
&\sum_{m,m'} {D^m}_{m'}(h') \ket{j,m}'_t\bra{j,m'}'_t \nonumber \\
         &= \sum_{m,m',m_s,m_t} U_{t,m_t m} {D^m}_{m'}(h') U^{\dagger}_{s,m'm_s} \ket{j,m_t}_t\bra{j,m_s}_s \nonumber \\ 
         & = \sum_{m_s,m_t} {D^{m_t}}_{m_s}(h) \ket{j,m_t}_t\bra{j,m_s}_s,
\end{align}
so that the following transformation property holds:
\begin{equation}
\hat{h} \rightarrow \hat{h}' = \hat{U}_t^{\dagger} \hat{h} 
\hat{U}_s. 
\label{HolonomyTransform}
\end{equation}

The expression (\ref{HolonomyTransform}) resembles the transformation 
rule of the holonomy under the local $SU(2)$ gauge transformation $\hat{U}$, 
which is $\hat{h} \rightarrow \hat{h}' = \hat{U}_s^{\dagger} \hat{h} 
\hat{U}_t$.  As shown in Ref. \cite{Mielczarek:2019srn}, the gauge 
transformation corresponds exactly to the change of bases at the source 
and the target, while the source and the target spaces are interchanged
in Eq. (\ref{HolonomyDef}). Actually, the choices of the order 
of the source and target spaces in Eq. (\ref{HolonomyDef}) is a convention, 
related by the hermitian conjugation. Consequently, while considering the 
hermitian conjugation of Eq. (\ref{HolonomyDef}), the transformation 
rule (\ref{HolonomyTransform}) takes the form usually expected in 
the case of holonomies, \emph{i.e.}:
\begin{equation}
\hat{h}^{\dagger}  \rightarrow \hat{h}^{\dagger'} = \hat{U}_s^{\dagger} \hat{h}^{\dagger} 
\hat{U}_t. 
\end{equation}

The transformation property leads to the concept of 
Wilson loops, which are gauge-invariant object 
defined as $W := {\rm tr} (h)$, where the source and
target spaces overlap. 

The space of holonomies, equipped with the Haar 
measure, forms a Hilbert space on $SU(2)$, \emph{i.e.}:
$\mathcal{H} = L^2(SU(2))$. Following the Peter-Weyl theorem,
one can express the space as the following direct sum:
\begin{equation}
L^2(SU(2)) = \oplus_j(V_j \otimes V_j^*).
\end{equation}
Consequently, the orthonormal basis states in the 
Hilbert space can be written as:
\begin{equation}
\hat{\varphi}(h)^j = \frac{1}{\sqrt{d_j}} \sum_{m,m'} {D^m}_{m'}(h) \ket{j,m}_t \bra{j,m'}_s \in V_j \otimes V_j^*. 
\label{SU2holonomystates}
\end{equation}

\begin{figure}[ht!]
    \centering
    \includegraphics[width=1\linewidth]{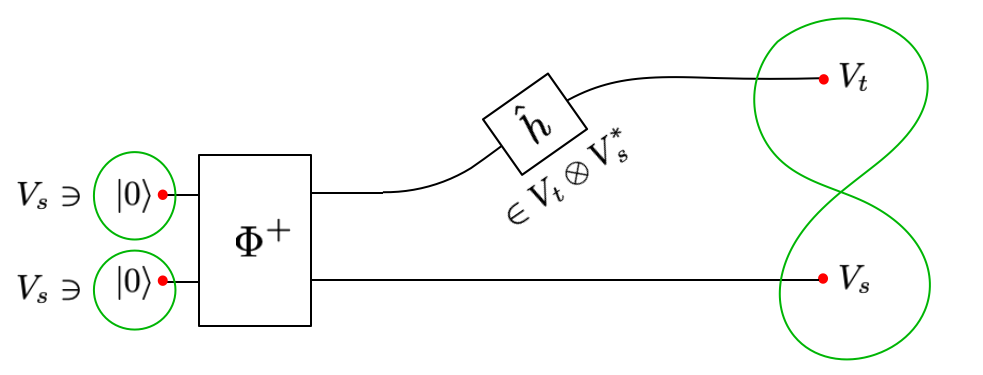}
    \caption{Diagrammatic illustration of the map (Choi-Jamio{\l}kowski isomorphism) defined in Eq. \ref{IsomorphismScalar}. Here, $V_s$ is the source Hilbert space, and $V_t$ is the target Hilbert space. The $\Phi^{+}$ box represents the quantum circuit (unitary operation) generating the $\ket{\Phi^{+}}$ state from a product state $\ket{0}\otimes \ket{0} \in V_s \otimes V_s$ (here, $\ket{0}$ are some initial states). According to the discussion 
    in Ref. \cite{Mielczarek:2019srn}, the final state is maximally entangled (which 
    comes from the unitarity of $\hat{h}$). For $j=1/2$ and $\ket{0}=\ket{1/2,-1/2}$ the $\Phi^{+}$ 
    corresponds to the quantum operator $\hat{\Phi}^{+} =\widehat{\rm CNOT} (\hat{H}\otimes\hat{\mathbb{I}})$, which involves Hadamard ($H$) and CNOT gates.}
    \label{fig:CJMap}
\end{figure}

An important observation is that states belonging to $V_j \otimes V_j^*$ can 
always be mapped into states in $V_j \otimes V_j$, which comes from the 
isomorphism between the space $V_j$ and its dual $V_j^*$. The isomorphism
is, however, not unique. The mapping between the bra and ket states 
can employ two possible bilinear forms, which are present in the case 
of the $SU(2)$ group. The first is the standard scalar product, which 
leads to $\langle jm|jm' \rangle =\delta^{m,m'}$. Here, the mapping can 
be  performed utilizing the state:
\begin{equation}
\ket{\Phi^{+}} = \frac{1}{\sqrt{d_j}} \sum_{m,m'} \delta^{m,m'} \ket{j,m}\ket{j,m'},
\end{equation}
so that the state dual to (\ref{SU2holonomystates}), is obtained via:
\begin{equation}
\ket{\varphi(h)^j} = (\hat{h}  \otimes \hat{\mathbb{I}})\ket{\Phi^{+}}. 
\label{IsomorphismScalar}
\end{equation}
A diagrammatic illustrator of the map is shown in Fig. \ref{fig:CJMap}.

This case has been studied in Ref. \cite{Mielczarek:2019srn}, where it 
has also been proven that by involving complex conjugation operation 
(which is antiunitary), the mapping can be constructed in agreement with 
the holonomy transformation rule. Furthermore, it is worth emphasizing 
that the expression (\ref{IsomorphismScalar}) is a special case of the 
seminal Choi-Jamio{\l}kowski isomorphism \cite{CHOI1975285,JAMIOLKOWSKI1972275},
playing an important role in quantum information. In the general case, the 
unitary map $\hat{\varphi}(h)^j$ can be replaced by an arbitrary quantum 
channel $\mathcal{E}$, which is a linear, completely positive, and
trace-preserving (CPTP) map:
\begin{equation}
\mathcal{E}: \mathcal{L}(V_s) \rightarrow  \mathcal{L}(V_t),
\end{equation}
where $\mathcal{L}(V)$ represents the space of the density  
operators over the Hilbert space $V$. 

The second bilinear form for the $SU(2)$ group is antisymmetric, and 
involves contraction with the antisymmetric tensor:
\begin{equation}
    g_{mm'}=\delta_{m,-m'}(-1)^{j-m}, 
    \label{deltammprime}
\end{equation}
so that $(jm|jm') =(-1)^{j-m} \delta_{m,-m'}$. The $g_{mm'}$ allows 
for raising and lowering the indices of the $SU(2)$ spinors, which
can be associated with the transition between the $V_j$ and $V_j^*$ spaces. 
Here, the mapping can be performed utilizing the singlet state:
\begin{equation}
\ket{\Psi^{-}} = \frac{1}{\sqrt{d_j}} \sum_{m,m'} g^{m,m'} \ket{j,m}\ket{j,m'},
\label{PsiState}
\end{equation}
so that the state dual to (\ref{SU2holonomystates}), in obtained via:
\begin{equation}
\ket{\varphi(h)^j} = (\hat{h}  \otimes \hat{\mathbb{I}})\ket{\Psi^{-}}, 
\label{IsomorphismAnty}
\end{equation}
so that: 
\begin{align}
\ket{\varphi(h)^j} &= \frac{1}{\sqrt{d_j}} \sum_{m,m'} (-1)^{j-m'} D^{m,m'}(h) \ket{j,m}_t \ket{j,-m'}_s  \nonumber \\
&\in V_j \otimes V_j, 
\label{holonomyiso}
\end{align}
which acts as follows:
\begin{equation}
\hat{\varphi}(h)^j  \in V_j \otimes V_j^* \rightarrow \ket{\varphi(h)^j}  \in V_j \otimes V_j. 
\label{isodef}
\end{equation}

The isomorphism given by Eq. \ref{holonomyiso} has been studied, e.g., 
in Ref. \cite{bianchi2023loop}. Importantly, two types of isomorphisms, 
employing the states $\ket{\Phi^{+}}$ and $\ket{\Psi^{-}}$, both lead to maximally 
entangled states, which can be shown by employing the unitarity of the 
$ {D^m}_{m'}(h)$ matrices. Since the second type of isomorphism has been 
more broadly considered in the LQG literature, we will refer to this case 
in what follows. 

As an illustrative example of the above discussion, let us consider 
the identity holonomy $h=\mathbb{I}$ in the fundamental representation, 
$j=1/2$, so that $\hat{h}=\begin{pmatrix}
1 & 0\\
0 & 1 \\
\end{pmatrix}$. Consequently, the state  $\hat{\varphi}(\mathbb{I})^{1/2} 
\in V_{1/2} \otimes V_{1/2}^*$  is $ \hat{\varphi}(\mathbb{I})^{1/2} = \frac{1}{\sqrt{2}} \begin{pmatrix}
1 & 0\\
0 & 1 \\
\end{pmatrix}$, so that its norm $||\hat{\varphi}(\mathbb{I})^{1/2}  || = \text{tr}
\sqrt{\hat{\varphi}(\mathbb{I})^{1/2}\hat{\varphi}(\mathbb{I})^{1/2,\dagger}} = 1$. 
The state (\ref{PsiState}) takes the form 
\begin{equation}
\ket{\Psi^{-}}= \frac{1}{\sqrt{2}}\left(\ket{\frac{1}{2},\frac{1}{2}}\ket{\frac{1}{2},-\frac{1}{2}}- \ket{\frac{1}{2},-\frac{1}{2}}\ket{\frac{1}{2},\frac{1}{2}} \right).
\end{equation}
In consequence,  the isomorphism (\ref{IsomorphismAnty}), leads to 
the holonomy state:
\begin{equation}
\ket{\varphi(\mathbb{I})}= \frac{1}{\sqrt{2}}\left(\ket{\frac{1}{2},\frac{1}{2}}\ket{\frac{1}{2},-\frac{1}{2}}- \ket{\frac{1}{2},-\frac{1}{2}}\ket{\frac{1}{2},\frac{1}{2}} \right). 
\end{equation}

\section{Spin network as tensor network}
\label{Sec.BasicDictionary}

Tensor networks are a versatile and powerful tool in both mathematics 
and computation, widely used to describe and analyze complex quantum 
systems, especially in quantum information theory and many-body physics. 
By offering an efficient framework for representing and manipulating 
large quantum states, they enable the study of otherwise intractable 
problems. Tensor networks play a pivotal role across several domains, 
including quantum field theory, condensed matter physics, and quantum 
information science, driving significant advances and deepening our 
understanding of these fields.

In Loop Quantum Gravity (LQG), the geometry of space is modeled as a 
quantum system with numerous degrees of freedom and a potentially 
complex structure. As a result, using tensor networks to study spin 
networks presents a natural and effective approach. To facilitate 
this analysis, our investigation will begin by developing a glossary 
that maps the corresponding components between spin networks and 
tensor networks.

\begin{figure}[ht!]
    \includegraphics[]{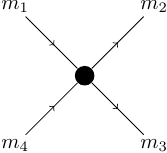}
    % \centering
    % \begin{tikzpicture}
    % \coordinate (a) at (-1,1);
    % \coordinate (b) at (1,1);
    % \coordinate (c) at (1,-1);
    % \coordinate (d) at (-1,-1);
    % \draw node[circle, fill] (n) {};
    % \draw[ingoing] (n) -- (a) node[pos=1.2] {$m_1$};
    % \draw[outgoing] (n) -- (b) node[pos=1.2] {$m_2$};
    % \draw[outgoing] (n) -- (c) node[pos=1.2] {$m_3$};
    % \draw[ingoing] (n) -- (d) node[pos=1.2] {$m_4$};
    % \end{tikzpicture}
    \caption{Single node of a spin network with two in-going and 
    two out-going links.}
    \label{fig:node_duud}
\end{figure}

The state of a 4-valent node in the spin networks framework is 
represented by a tensor with four magnetic indices ($m_1,m_2,m_3,m_4$) 
which correspond to links connected to this node and one more index 
($k$) corresponding to the internal degree of freedom of the node.
The tensor is called an intertwiner: 
\begin{equation}
    \iota_{k}^{m_1m_2m_3m_4}.
\end{equation}
In the spin network notation, the upper and lower indices correspond 
to outgoing and ingoing links, respectively. For instance: 
\begin{equation}
    \iota\indices{_{(k)m_1}^{m_2m_3}_{m_4}}\in 
    {\rm Inv}_{SU(2)}\left(V^*\otimes V\otimes V\otimes V^*\right),    
\end{equation}
where $V$ is the Hilbert space of a single spin, and $V^*$ is a 
dual space to $V$. This tensor can be represented by a spin network 
diagram, as shown in Fig. \ref{fig:node_duud}, where the index $k$ 
has not been explicitly marked. 

\begin{figure}[ht!]
    \includegraphics[]{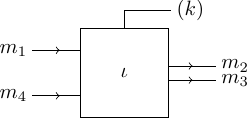}
    % \centering
    % \begin{tikzpicture}[
    % tensor/.style={rectangle, draw, minimum size=15mm},
    % ]
    % \coordinate (a) at (0.8,0);
    % \coordinate (b) at (0.8,0);
    
    % \node[tensor] (i) {$\iota$};
    % \draw[ingoing] (i.-207) -- ++(-0.8,0) node[pos=1.4] {$m_1$};
    % \draw[outgoing] (i.9) -- ++(0.8,0) node[pos=1.4] {$m_2$};
    % \draw[outgoing] (i.-9) -- ++(0.8,0) node[pos=1.4] {$m_3$};
    % \draw[ingoing] (i.207) -- ++(-0.8,0) node[pos=1.4] {$m_4$};
    % \draw (i.north) -- ++(0,0.3) -- ++(0.8,0) node[pos=1.4] {$(k)$};
    % \end{tikzpicture}
    \caption{Single node with two in-going and two out-going links.}
    \label{fig:node_tensor_duud}
\end{figure}

The same node can also be represented by employing a tensor network 
diagram. A rectangle can be drawn, with four legs corresponding to 
the indices $m_1, m_2, m_3$, and $m_4$, while the fifth leg corresponds 
to the index $k$. The positions of the indices are indicated 
by arrows, with the outgoing arrow corresponding to the upper 
index and the ingoing arrow corresponding to the lower index, 
as shown in Fig. \ref{fig:node_tensor_duud}.

\begin{figure}[ht!]
    \includegraphics[]{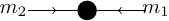}
    % \centering
    % \begin{tikzpicture}
    % \draw node[circle, fill] (n) {};
    % \draw[ingoing] (n) -- ++(1,0) node[pos=1.2] {$m_1$};
    % \draw[ingoing] (n) -- ++(-1,0) node[pos=1.3] {$m_2$};
    % \end{tikzpicture}
    \caption{Bivalent node.}
    \label{fig:bivalent}
\end{figure}

In the preceding paragraph, the Wigner $4j$-symbol is denoted by the symbol $\iota$ and is defined as follows:
\begin{align}
    &\iota_{k}^{m_1m_2m_3m_4} = \nonumber \\
    &= \sqrt{2k+1}\left(\begin{array}{ccc}
         j_1 & j_2 & k \\
         m_1& m_2 & m \\
    \end{array}\right)
    g_{mm'}
    \left(\begin{array}{ccc}
         k & j_3 & j_4 \\
         m'& m_3 & m_4 \\
    \end{array}\right),
\end{align}
where $g_{mm'}$ is defined in Eq.  \ref{deltammprime} and provides an 
isomorphism between $V$ and its dual space $V^*$. Employing the 
$g_{mm'}$ tensor, the indices of the Wigner $4j$-symbols may be lowered 
or raised. In the language of spin networks, the tensor $g_{mm'}$ may 
be regarded as a bivalent node possessing two incoming links and represented 
as shown in Fig. \ref{fig:bivalent}.

We can omit the dot symbol as bivalent nodes lack internal degrees 
of freedom. In the case of tensor network representation, the tensor 
will be depicted as a circle with two ingoing links as shown in 
Fig.\ref{fig:g_dd}:

\begin{figure}[ht!]
    \includegraphics[]{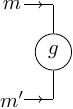}
    % \centering
    % \begin{tikzpicture}[
    % vector/.style={circle, draw},
    % ]
    % \coordinate (a) at (0,0.8);
    % \coordinate (b) at (-0.5,0.8);
    % \coordinate (c) at (0,-0.8);
    % \coordinate (d) at (-0.5,-0.8);
    % \node[vector] (g) {$g$};
    % \draw (g.north) -- (a);
    % \draw (g.south) -- (c);
    % \draw[ingoing] (a) -- (b) node[pos=1.4] {$m$};
    % \draw[ingoing] (c) -- (d) node[pos=1.4] {$m'$};
    % \end{tikzpicture}
    \caption{Tensor $g_{mm'}$.}
    \label{fig:g_dd}
\end{figure}

\begin{figure}[ht!]
    \includegraphics[]{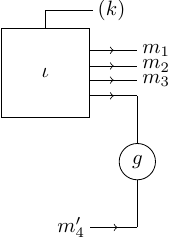}
    % \centering
    % \begin{tikzpicture}[
    % tensor/.style={rectangle, draw, minimum size=15mm},
    % vector/.style={circle, draw},
    % ]
    % \coordinate (a) at (0.8,0);
    % \coordinate (b) at (0.8,0);
    
    % \node[tensor] (i) {$\iota$};
    % \draw[outgoing] (i.27) -- ++(0.8,0) node[pos=1.4] {$m_1$};
    % \draw[outgoing] (i.9) -- ++(0.8,0) node[pos=1.4] {$m_2$};
    % \draw[outgoing] (i.-9) -- ++(0.8,0) node[pos=1.4] {$m_3$};
    % \draw[outgoing] (i.-27) -- ++(0.8,0);
    % \draw ($(i.-27)+(0.8,0)$)-- ++(0,-0.8) node[vector, below] (g) {$g$};
    % \draw (g.south) -- ++(0,-0.8);
    % \draw[ingoing] ($(g.south)+(0,-0.8)$)-- ++(-0.8,0) node[pos=1.4] {$m_4'$};
    
    % \draw (i.north) -- ++(0,0.3) -- ++(0.8,0) node[pos=1.4] {$(k)$};
    % \end{tikzpicture}
    \caption{Lowering index of Wigner $4j$-symbol.}
    \label{fig:node_tensor_uuuu_to_uuud}
\end{figure}

As we already said, the $g$ tensor can be used to change the position 
of magnetic indices, for example:
\begin{equation}    \iota\indices{_{(k)}^{m_1m_2m_3}_{m_4}}=g_{m_4m_4'}\iota^{m_1m_2m_3m_4'}_k.
\end{equation}
We can represent this operation using a tensor network, as shown in Fig. 
\ref{fig:node_tensor_uuuu_to_uuud}.

The links of spin networks are labeled by $SU(2)$ holonomies, which 
can be represented as tensors using the Wigner D-matrices:
\begin{equation}
    D(h)\indices{^{m'}_m},
\end{equation}
where $h\in SU(2)$.

\begin{figure}[ht!]
    \includegraphics[]{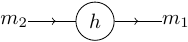}
    % \centering
    % \begin{tikzpicture}[
    % vector/.style={circle, draw},
    % ]
    % \node[vector] (v) {$h$};
    % \draw[outgoing] (v.east) -- ++(0.8,0) node[pos=1.3] {$m_1$};
    % \draw[ingoing] (v.west) -- ++(-0.8,0) node[pos=1.3] {$m_2$};
    % \end{tikzpicture}
    \caption{Tensor representing holonomy $D\left(h\right)\indices{^{m_1}_{m_2}}$.}
    \label{fig:holonomy_tensor}
\end{figure}

In the spin network diagrams, holonomies are represented solely as 
labels of links. However, in the tensor network diagrams, we will 
employ circles with an indication of the holonomy represented as 
$h$, as depicted in Fig. \ref{fig:holonomy_tensor}.

\begin{figure}[ht!]
    \includegraphics[]{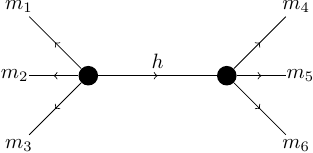}
    % \centering
    % \begin{tikzpicture}
    % \draw node[circle, fill] (n) {};
    % \draw[outgoing] (n) -- ++(-1,1) node[pos=1.2] {$m_1$};
    % \draw[outgoing] (n) -- ++(-1,0) node[pos=1.3] {$m_2$};
    % \draw[outgoing] (n) -- ++(-1,-1) node[pos=1.2] {$m_3$};
    % \draw[outgoing] (n.east) -- ++(2,0) node[pos=0.5, above] {$h$} node[circle, fill, right] (n2) {};
    % \draw[outgoing] (n2) -- ++(1,1) node[pos=1.2] {$m_4$};
    % \draw[outgoing] (n2) -- ++(1,0) node[pos=1.3] {$m_5$};
    % \draw[outgoing] (n2) -- ++(1,-1) node[pos=1.2] {$m_6$};
    % \end{tikzpicture}
    \caption{Part of the spin network corresponding to Eq. \ref{PartofSN}. }
    \label{fig:two_nodes}
\end{figure}

Let us consider a part of the spin network shown Fig.~\ref{fig:two_nodes} corresponding to the expression:
\begin{equation}
    \begin{split}
        &\iota^{m_1m_2m_3m}D\left(h\right)\indices{^{m'}_{m}}\iota\indices{_{m'}^{m_4m_5m_6}}
        =\\&
        \iota\indices{^{m_1m_2m_3m}}D\left(h\right)\indices{^{m'}_{m}}g_{m'm''}\iota\indices{^{m''m_4m_5m_6}}
    \end{split}
    \label{PartofSN}
\end{equation}

\begin{figure}[ht!]
    \includegraphics[]{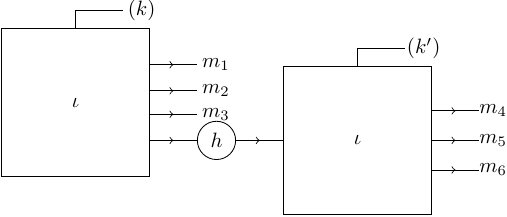}
    % \centering
    % \begin{tikzpicture}[
    % tensor/.style={rectangle, draw, minimum size=25mm},
    % vector/.style={circle, draw},
    % ]
    % \coordinate (a) at (0.8,0);
    % \coordinate (b) at (0.8,0);
    
    % \node[tensor] (i) {$\iota$};
    % \draw (i.north) -- ++(0,0.3) -- ++(0.8,0) node[pos=1.4] {$(k)$};
    % \draw[outgoing] (i.27) -- ++(0.8,0) node[pos=1.4] {$m_1$};
    % \draw[outgoing] (i.9) -- ++(0.8,0) node[pos=1.4] {$m_2$};
    % \draw[outgoing] (i.-9) -- ++(0.8,0) node[pos=1.4] {$m_3$};
    % \draw[outgoing] (i.-27) -- ++(0.8,0) node[vector, right] (h) {$h$};
    % \draw[outgoing] (h.east) -- ++(0.8,0) node[tensor, right] (i2) {$\iota$};
    % \draw (i2.north) -- ++(0,0.3) -- ++(0.8,0) node[pos=1.4] {$(k')$};
    % \draw[outgoing] ($(i2.east)+(0,0.5)$) -- ++(0.8,0) node[pos=1.3] {$m_4$};
    % \draw[outgoing] (i2.east) -- ++(0.8,0) node[pos=1.3] {$m_5$};
    % \draw[outgoing] ($(i2.east)+(0,-0.5)$) -- ++(0.8,0) node[pos=1.3] {$m_6$};
    % \end{tikzpicture}
    \caption{Tensor network representation of $\iota^{m_1m_2m_3m}D\left(h\right)\indices{^{m'}_{m}}\iota\indices{_{m'}^{m_4m_5m_6}}$}
    \label{fig:two_nodes_tensor}
\end{figure}

The spin network can be represented using two 
$\iota$ tensors with properly oriented legs and a holonomy tensor 
$h$, as depicted in the Fig.~\ref{fig:two_nodes_tensor}.

Utilizing the building blocks outlined in this section makes 
it feasible to create arbitrary states of spin networks as 
tensor networks.

\section{Quantum circuits construction}
\label{Sec.QuantumCircuits}

As described in the previous section, we have created a dictionary 
that outlines the relationship between spin networks and tensor networks. 
This dictionary serves as a means of translating spin networks into 
quantum circuits, and tensor networks are used as an intermediate 
step in this process.

In this study, we concentrate on spin networks with fixed spin labels, 
corresponding to the fundamental representation of $SU(2)$, $j=1/2$. 
Each tensor leg is two-dimensional and can be straightforwardly represented 
with a qubit. Additionally, every tensor can be presented as a quantum gate, 
employing ancillary qubits or measurements for non-unitary tensors. Also, 
it is convenient that each node in this scenario possesses two internal 
degrees of freedom; hence, its state can be expressed with a single qubit.

First, we prepare a quantum circuit that represents the 
tensor $\iota_{(k)}^{m_1m_2m_3m_4}$. The corresponding tensor 
network is a rectangle with four $m_i$ legs and a single $k$ leg. 
One requires a 5-qubit gate to act on the fixed initial state 
(e.g., $\ket{00000}$) since the quantum gate is a unitary operator 
that needs the same number of incoming and outgoing qubits. 
The choice of the initial state is a matter of convention, and 
in what follows, we fix it to be $\ket{00000}$. However, other 
choices can also be considered, resulting in a different expression 
for the corresponding quantum circuit. 

The five outgoing qubits correspond to indices $k,m_1,m_2,m_3,m_4$. 
A schematic form of this gate is presented in Fig. \ref{fig:A}, 
whereas the specific form in terms of simple gates is included in 
Appendix A.

\begin{figure}[ht!]
    \includegraphics[scale=1]{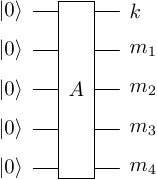}
    % \leavevmode
    % \centering
    % \Qcircuit @C=1.2em @R=1em {
    %     \lstick{\ket{0}}& \multigate{4}{A} & \rstick{k} \qw \\
    %     \lstick{\ket{0}}& \ghost{A} & \rstick{m_1} \qw\\
    %     \lstick{\ket{0}}& \ghost{A} & \rstick{m_2} \qw\\
    %     \lstick{\ket{0}}& \ghost{A} & \rstick{m_3} \qw\\
    %     \lstick{\ket{0}}& \ghost{A} & \rstick{m_4} \qw
    %     }
    \caption{Quantum circuit for $\iota_{(k)}^{m_1m_2m_3m_4}$.}
    \label{fig:A}
\end{figure}

\begin{figure}[ht!]
    \includegraphics[]{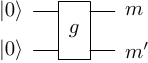}
    % \leavevmode
    % \centering
    % \Qcircuit @C=1.2em @R=1em {
    %     \lstick{\ket{0}}& \multigate{1}{g} & \rstick{m} \qw \\
    %     \lstick{\ket{0}}& \ghost{g} & \rstick{m'} \qw\\
    %     }
    \caption{Tensor representation of $g^{mm'}$.}
    \label{fig:g}
\end{figure}

\begin{figure}[ht!]
    \includegraphics[]{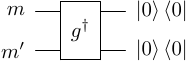}
    % \leavevmode
    % \centering
    % \Qcircuit @C=1.2em @R=1em {
    %     \lstick{m} & \multigate{1}{g^{\dagger}} & \rstick{\ket{0}\bra{0}} \qw \\
    %     \lstick{m'} & \ghost{g^{\dagger}} & \rstick{\ket{0}\bra{0}} \qw\\
    %     }
    \caption{Tensor representation of $g_{mm'}$.}
    \label{fig:g_dd_circ}
\end{figure}

\begin{figure}[ht!]
    \includegraphics[]{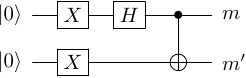}
    % \leavevmode
    % \centering
    % \Qcircuit @C=1.2em @R=1em {
    %     \lstick{\ket{0}}& \gate{X} & \gate{H} & \ctrl{1} & \rstick{m} \qw \\
    %     \lstick{\ket{0}}& \gate{X} & \qw & \targ & \rstick{m'} \qw\\
    %     }
    \caption{Quantum circuit for $g^{mm'}$.}
    \label{fig:g_explicit}
\end{figure}

\begin{figure}[ht!]
    \includegraphics[]{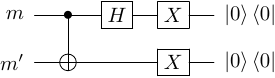}
    % \leavevmode
    % \centering
    % \Qcircuit @C=1.2em @R=1em {
    %     \lstick{m}&   \ctrl{1} &\gate{H} &\gate{X} & \rstick{\ket{0}\bra{0}} \qw \\
    %     \lstick{m'}& \targ &  \qw & \gate{X} &\rstick{\ket{0}\bra{0}} \qw\\
    %     }
    \caption{Quantum circuit for $g_{mm'}$.}
    \label{fig:g_dd_explicit}
\end{figure}

As the next step, we proceed to construct the tensor $g$. To obtain $g^{mm'}$ 
with upper indices, we require a quantum gate that acts on two fixed qubits 
and two outgoing qubits that represent $m$ and $m'$, respectively (as shown 
in Fig. \ref{fig:g}). On the other hand, for $g_{mm'}$, a different approach 
is necessary. We need a gate that operates on two qubits, representing $m$ 
and $m'$, and two fixed outgoing qubits. This can be accomplished by projecting 
onto a fixed state, such as $\ket{00}$, illustrated in Fig. \ref{fig:g_dd_circ}. 
These two cases are depicted as explicit circuits in Figs. \ref{fig:g_explicit} 
and \ref{fig:g_dd_explicit}, respectively. In the aforementioned construction, 
the ingoing/outgoing edges of the tensor network correspond to the ingoing/outgoing 
qubits of the quantum gate. 

\begin{figure}[ht!]
    \includegraphics[]{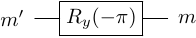}
    % \leavevmode
    % \centering
    % \Qcircuit @C=1.2em @R=1em {
    %     \lstick{m'}& \gate{R_y(-\pi)} & \rstick{m} \qw \\
    %     }
    \caption{Quantum circuit for $g^{m}_{m'}$, where $R_y(\theta)$ corresponds to 
    the rotation operator.}
    \label{fig:g_ud_explicit}
\end{figure}

\begin{figure}[ht!]
    \includegraphics[]{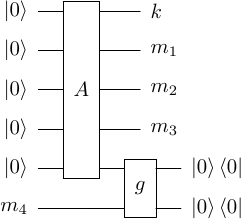}
    % \leavevmode
    % \centering
    % \Qcircuit @C=1.2em @R=1em {
    %     \lstick{\ket{0}}& \multigate{4}{A} & \rstick{k} \qw & \\
    %     \lstick{\ket{0}}& \ghost{A}  & \rstick{m_1} \qw & \\
    %     \lstick{\ket{0}}& \ghost{A}  &  \rstick{m_2} \qw & \\
    %     \lstick{\ket{0}}& \ghost{A}  & \rstick{m_3} \qw & \\
    %     \lstick{\ket{0}}& \ghost{A} & \multigate{1}{g} &\rstick{\ket{0}\bra{0}} \qw\\
    %     \lstick{m_4} & \qw & \ghost{g}  & \rstick{\ket{0}\bra{0}} \qw\\
    %     }
    \caption{Lowering an index in terms of a quantum circuit.}
    \label{fig:A_uuud}
\end{figure}

For $g^{m}_{m'}$, we may utilize a single qubit to construct a quantum gate, 
requiring only one ingoing and one outgoing link. As seen in Fig. \ref{fig:g_ud_explicit}, a single qubit rotation is sufficient.
Now, let us move to a node with a different index position. To determine 
the quantum circuit for $\iota\indices{_{(k)}^{m_1m_2m_3}_{m_4}}$, we can 
use the circuit for $\iota\indices{_{(k)}^{m_1m_2m_3m_4}}$ and for $g^{mm'}$.
We can combine them according to the formula 
\begin{equation}
    \iota\indices{_{(k)}^{m_1m_2m_3}_{m_4}}=\iota_{(k)}^{m_1m_2m_3m_4'}g_{m_4'm_4}
\end{equation}
To implement this
method, we require an ancilla qubit as shown in Figure \ref{fig:A_uuud}.

\begin{figure}[ht!]
    \includegraphics[]{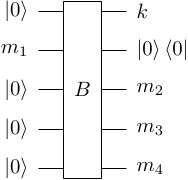}
    % \leavevmode
    % \centering
    % \Qcircuit @C=1.2em @R=1em {
    %     \lstick{\ket{0}}& \multigate{4}{B} & \rstick{k} \qw \\
    %     \lstick{m_1}& \ghost{B} & \rstick{\ket{0}\bra{0}} \qw\\
    %     \lstick{\ket{0}}& \ghost{B} & \rstick{m_2} \qw\\
    %     \lstick{\ket{0}}& \ghost{B} & \rstick{m_3} \qw\\
    %     \lstick{\ket{0}}& \ghost{B} & \rstick{m_4} \qw
    %     }
    \caption{Quantum circuit for $\iota\indices{_{(k)}^{m_1m_2m_3}_{m_4}}$.}
    \label{fig:B}
\end{figure}

\begin{figure}[ht!]
    \includegraphics[]{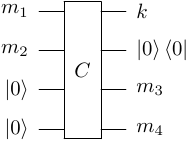}
    % \leavevmode
    % \centering
    % \Qcircuit @C=1.2em @R=1em {
    %     \lstick{m_1}& \multigate{3}{C} & \rstick{k} \qw \\
    %     \lstick{m_2}& \ghost{C} & \rstick{\ket{0}\bra{0}} \qw\\
    %     \lstick{\ket{0}}& \ghost{C} & \rstick{m_3} \qw\\
    %     \lstick{\ket{0}}& \ghost{C} & \rstick{m_4} \qw\\
    %     }
    \caption{Quantum circuit for $\iota\indices{_{(k)}_{m_1m_2}^{m_3m_4}}$.}
    \label{fig:C}
\end{figure}

\begin{figure}[ht!]
    \includegraphics[]{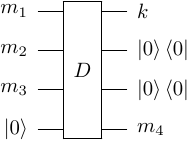}
    % \leavevmode
    % \centering
    % \Qcircuit @C=1.2em @R=1em {
    %     \lstick{m_1}& \multigate{3}{D} & \rstick{k} \qw \\
    %     \lstick{m_2}& \ghost{D} & \rstick{\ket{0}\bra{0}} \qw\\
    %     \lstick{m_3}& \ghost{D} & \rstick{\ket{0}\bra{0}} \qw\\
    %     \lstick{\ket{0}}& \ghost{D} & \rstick{m_4} \qw\\
    %     }
    \caption{Quantum circuit for $\iota\indices{_{(k)}_{m_1m_2m_3}^{m_4}}$.}
    \label{fig:D}
\end{figure}

\begin{figure}[ht!]
    \includegraphics[]{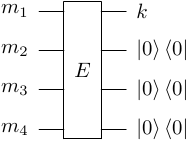}
    % \leavevmode
    % \centering
    % \Qcircuit @C=1.2em @R=1em {
    %     \lstick{m_1}& \multigate{3}{E} & \rstick{k} \qw \\
    %     \lstick{m_2}& \ghost{E} & \rstick{\ket{0}\bra{0}} \qw\\
    %     \lstick{m_3}& \ghost{E} & \rstick{\ket{0}\bra{0}} \qw\\
    %     \lstick{m_4}& \ghost{E} & \rstick{\ket{0}\bra{0}} \qw\\
    %     }
    \caption{Quantum circuit for $\iota\indices{_{(k)}_{m_1m_2m_3m_4}}$.}
    \label{fig:E}
\end{figure}

However, there is an alternative approach. We can create five-qubit 
circuits that operate on four fixed qubits and one qubit representing 
the incoming link. The outcomes of this gate are expressed on three 
qubits representing outgoing links, one qubit representing the index 
$k$, and one fixed qubit using projection, as depicted in Fig. \ref{fig:B}. 
We also require three additional circuits that can accommodate various 
configurations of upper and lower indices, as illustrated in Figs. \ref{fig:C}, 
\ref{fig:D}, and \ref{fig:E}.


The explicit forms of all these circuits, expressed in terms 
of simple gates can be found in Appendix A.

The subsequent stage in constructing a quantum circuit representing a 
spin network is to incorporate a gate that represents the holonomy tensor, 
denoted by $h$. In the case of an Ising-type spin network, we must apply 
a single qubit gate that corresponds to the Wigner matrix of the given 
holonomy. Please see the diagram below, Fig. \ref{fig:h}:
\begin{figure}[ht!]
    \includegraphics[]{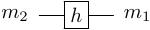}
    % \leavevmode
    % \centering
    % \Qcircuit @C=1.2em @R=1em {
    %     \lstick{m_2}& \gate{h} & \rstick{m_1} \qw \\
    %     }
    \caption{Quantum circuit for $D\left(h\right)\indices{^{m_1}_{m_2}}$.}
    \label{fig:h}
\end{figure}

In general, one could also consider superpositions over holonomies. 
Consequently, the relevant tensor would not be unitary and an ancilla 
qubit with measurement is necessary for non-unitary gate preparation.

\begin{figure}[ht!]
    \includegraphics[]{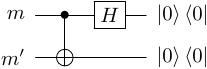}
    % \leavevmode
    % \centering
    % \Qcircuit @C=1em @R=1em {
    %     \lstick{m}& \ctrl{1} &\gate{H} &\rstick{\ket{0}\bra{0}} \qw \\
    %     \lstick{m'}& \targ &\qw  &\rstick{\ket{0}\bra{0}} \qw \\
    %     }
    \caption{Quantum circuit for $\mathbb{I}$.}
    \label{fig:id_circ}
\end{figure}

\begin{figure}[ht!]
    \includegraphics[]{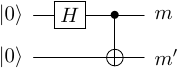}
    % \leavevmode
    % \centering
    % \Qcircuit @C=1em @R=1em {
    %     \lstick{\ket{0}}&\gate{H}&  \ctrl{1} &\rstick{m} \qw \\
    %     \lstick{\ket{0}}& \qw &\targ  &\rstick{m'} \qw \\
    %     }
    \caption{Quantum circuit for $\mathbb{I}$.}
    \label{fig:id_circ2}
\end{figure}

In certain cases, supplementary building blocks might be useful.  
For instance, a trivial tensor network with only one link relates 
to a bare quantum wire lacking a gate. However, if the link between 
qubits needs to be utilized, the circuit depicted in Figs.~\ref{fig:id_circ}, 
\ref{fig:id_circ2} is required.

\subsection{The isomorphism}

As already discussed in Sec. \ref{Sec.Holonomies} there is an 
isomorphism relating the holonomies, understood as unitary 
maps, and the maximally entangled states. Let us consider  
the associated tensor network and the quantum circuit here. 

\begin{figure}[ht!]
    \includegraphics[]{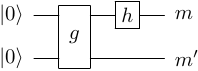}
     % \leavevmode
     % \centering
     % \Qcircuit @C=1.2em @R=1em {
     %     \lstick{\ket{0}}& \multigate{1}{g} &\gate{h}& \rstick{m} \qw \\
     %     \lstick{\ket{0}}& \ghost{g} &\qw& \rstick{m'} \qw\\
     %     }
     \caption{Quantum circuit for the isomorphism $D\left(h\right)\indices{^{m}_{m''}}g^{m''m'}=D\left(h\right)\indices{^{mm'}}$.}
     \label{fig:isomorphism}
\end{figure}

The core element of the isomorphism is the state (\ref{PsiState}), 
associated with the metric $g^{mm'}$. The metric raises the second 
index of $D\left(h\right)\indices{^{m}_{m'}}$, so that in consequence:
\begin{equation}
D\left(h\right)\indices{^{m}_{m''}}g^{m''m'}=D\left(h\right)\indices{^{mm'}}.
\end{equation}
Employing the circuits shown in Fig. \ref{fig:g} and Fig. \ref{fig:h} 
we obtain the circuit presented in Fig. \ref{fig:isomorphism}. 

Let us notice that the method presented in \cite{czelusta2023quantum} is actually 
a special case of the method presented here. The three-qubit-projection operator 
($\hat{W}^\dagger$ in \cite{czelusta2023quantum}) is equivalent to our operator $E$.  
The singlet states, which generate links in Ref. \cite{czelusta2023quantum}, 
are equivalent to $h=\mathbb{I}$. 

\subsection{Example: Pentagram}

\begin{figure}[ht!]
    \includegraphics[]{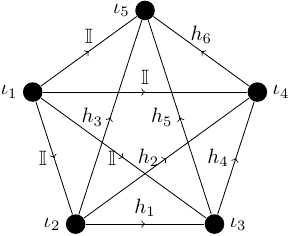}
    % \centering
    % \begin{tikzpicture}[
    % vector/.style={circle, draw},
    % ]
    % \draw (90+360/5:2) node[circle, fill] (n1) {} node[shift={(-0.4,0)}] {$\iota_1$};
    % \draw (90+360/5*2:2) node[circle, fill] (n2) {} node[shift={(-0.4,0)}] {$\iota_2$};
    % \draw (90+360/5*3:2) node[circle, fill] (n3) {} node[shift={(0.4,0)}] {$\iota_3$};
    % \draw (90+360/5*4:2) node[circle, fill] (n4) {} node[shift={(0.4,0)}] {$\iota_4$};
    % \draw (90+360/5*5:2) node[circle, fill] (n5) {} node[shift={(-0.4,0)}] {$\iota_5$};
    % \draw[outgoing] (n1) -- (n2) node[pos=0.5, left] {$\mathbb{I}$};
    % \draw[outgoing] (n1) -- (n3) node[pos=0.5, left] {$\mathbb{I}$};
    % \draw[outgoing] (n1) -- (n4) node[pos=0.5, above] {$\mathbb{I}$};
    % \draw[outgoing] (n1) -- (n5) node[pos=0.5, above] {$\mathbb{I}$};
    % \draw[outgoing] (n2) -- (n3) node[pos=0.5, above] {$h_1$};
    % \draw[outgoing] (n2) -- (n4) node[pos=0.5, left] {$h_2$};
    % \draw[outgoing] (n2) -- (n5) node[pos=0.5, left] {$h_3$};
    % \draw[outgoing] (n3) -- (n4) node[pos=0.5, left] {$h_4$};
    % \draw[outgoing] (n3) -- (n5) node[pos=0.5, left] {$h_5$};
    % \draw[outgoing] (n4) -- (n5) node[pos=0.5, above] {$h_6$};
    % \end{tikzpicture}
    \caption{Pentagram spin network.}
    \label{fig:pentagram}
\end{figure}

\begin{figure}[ht!]
    \includegraphics[]{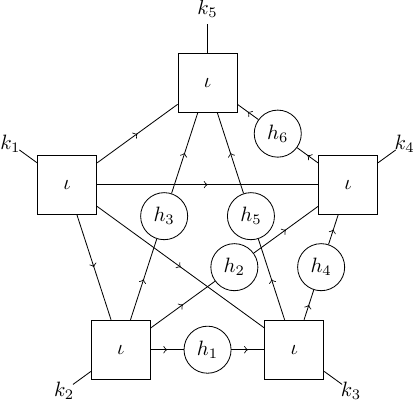}
    % \centering
    % \begin{tikzpicture}[
    % vector/.style={circle, draw},
    % tensor/.style={rectangle, draw, minimum size=10mm},
    % ]
    % \draw (90+360/5:2.5) node[tensor] (n1) {$\iota$};
    % \draw (90+360/5*2:2.5) node[tensor] (n2) {$\iota$};
    % \draw (90+360/5*3:2.5) node[tensor] (n3) {$\iota$};
    % \draw (90+360/5*4:2.5) node[tensor] (n4) {$\iota$};
    % \draw (90+360/5*5:2.5) node[tensor] (n5) {$\iota$};
    % \draw[outgoing] (n1) -- (n2);
    % \draw[outgoing] (n1) -- (n3);
    % \draw[outgoing] (n1) -- (n4);
    % \draw[outgoing] (n1) -- (n5);
    % \draw ($0.5*(n2)+0.5*(n3)$) node[vector] (h1) {$h_1$};
    % \draw[outgoing] (n2) -- (h1);
    % \draw[outgoing] (h1) -- (n3);
    % \draw ($0.5*(n2)+0.5*(n4)$) node[vector] (h2) {$h_2$};
    % \draw[outgoing] (n2) -- (h2);
    % \draw[outgoing] (h2) -- (n4);
    % \draw ($0.5*(n2)+0.5*(n5)$) node[vector] (h3) {$h_3$};
    % \draw[outgoing] (n2) -- (h3);
    % \draw[outgoing] (h3) -- (n5);
    % \draw ($0.5*(n3)+0.5*(n4)$) node[vector] (h4) {$h_4$};
    % \draw[outgoing] (n3) -- (h4);
    % \draw[outgoing] (h4) -- (n4);
    % \draw ($0.5*(n3)+0.5*(n5)$) node[vector] (h5) {$h_5$};
    % \draw[outgoing] (n3) -- (h5);
    % \draw[outgoing] (h5) -- (n5);
    % \draw ($0.5*(n4)+0.5*(n5)$) node[vector] (h6) {$h_6$};
    % \draw[outgoing] (n4) -- (h6);
    % \draw[outgoing] (h6) -- (n5);
    % \draw (n1) -- ++(144:1) node[pos=1.5] {$k_1$};
    % \draw (n2) -- ++(216:1) node[pos=1.5] {$k_2$};
    % \draw (n3) -- ++(-36:1) node[pos=1.5] {$k_3$};
    % \draw (n4) -- ++(36:1) node[pos=1.5] {$k_4$};
    % \draw (n5) -- ++(90:1) node[pos=1.5] {$k_5$};
    % \end{tikzpicture}
    \caption{Tensor network corresponding to pentagram}
    \label{fig:pentagram_tn}
\end{figure}

\begin{figure}[ht!]
    \includegraphics[scale=0.8]{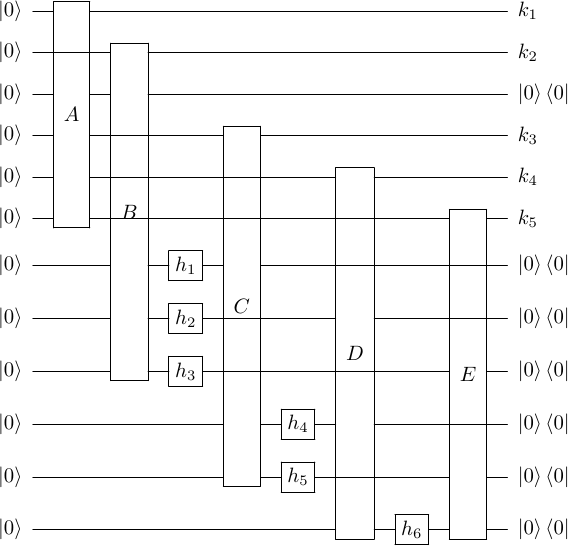}
        \caption{Quantum circuit for the pentagram spin network. Explicit 
        expressions of the multi-qubit gates $A$, $B$, $C$, $D$ and $E$ 
        can be found in Appendix A. }
        \label{fig:pentagramcirc}
\end{figure}

Let us examine a detailed construction of a pentagram in the form of 
spin and tensor networks, as presented in Figs.~\ref{fig:pentagram} 
and~\ref{fig:pentagram_tn}. Four links have been selected, each with 
holonomies that are trivial $\mathbb{I}$, which is always achievable 
using gauge invariance of the spin network. In order to create the 
circuit, one must first determine the order of the $\iota$ tensors and 
then assign gates $A$, $B$, $C$, $D$, and $E$ consecutively to these 
tensors. One must select the qubits upon which our gates will operate 
to mirror the topology of the spin network. Then, one applies $h$ gates 
between them if the link has a non-trivial holonomy. The diagram of 
the resulting circuit is presented in Fig. \ref{fig:pentagramcirc}.

\subsection{Contraction}

\begin{figure}[ht!]
    \includegraphics[]{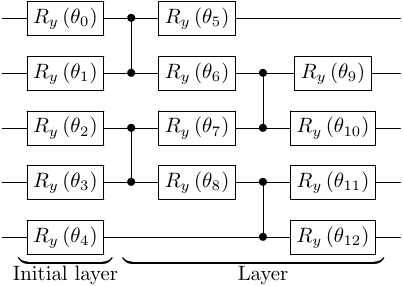}
        % \leavevmode
        % \centering
        % \Qcircuit @C=1.2em @R=1em {
        %     & \gate{R_y\left(\theta_0\right)} & \ctrl{1} & \gate{R_y\left(\theta_5\right)} & \qw & \qw & \qw \\
        %     & \gate{R_y\left(\theta_1\right)} & \ctrl{-1} & \gate{R_y\left(\theta_6\right)} &\ctrl{1}& \gate{R_y\left(\theta_9\right)}& \qw \\
        %     & \gate{R_y\left(\theta_2\right)} & \ctrl{1} & \gate{R_y\left(\theta_7\right)} &\ctrl{-1}& \gate{R_y\left(\theta_{10}\right)} & \qw \\
        %     & \gate{R_y\left(\theta_3\right)} & \ctrl{-1} & \gate{R_y\left(\theta_8\right)} &\ctrl{1} & \gate{R_y\left(\theta_{11}\right)} & \qw \\
        %     & \gate{R_y\left(\theta_4\right)} & \qw & \qw & \ctrl{-1} & \gate{R_y\left(\theta_{12}\right)} & \qw \\
        %     &\mbox{Initial layer} & & & \mbox{Layer} & &\\
        %     {\gategroup{5}{2}{5}{2}{0.8em}{_)}}
        %     {\gategroup{5}{3}{5}{6}{0.8em}{_)}}
        % }
        \caption{Simplified-two-design ansatz with one layer.}
        \label{fig:s2d}
    \end{figure}

Using the tensor network that represents the pentagram (as illustrated in Fig.~\ref{fig:pentagram_tn}), internal links can be contracted, and a five-legged 
tensor representing the pentagram state can be obtained. This procedure can 
be mirrored when using a quantum computer to transfer the obtained state to a 
parametrized ansatz. The ansatz shown in Fig.~\ref{fig:s2d} can be used and applied 
to qubits $0,5,2,3,4$ in Fig.~\ref{fig:pentagramcirc}.

One can adjust the ansatz parameters to represent the state of the 
pentagram through conjugated ansatz. This technique employs the same 
transfer method outlined in Ref. \cite{khatri2019quantum}.

\subsection{Open spin networks}

\begin{figure}[ht!]
    \includegraphics[]{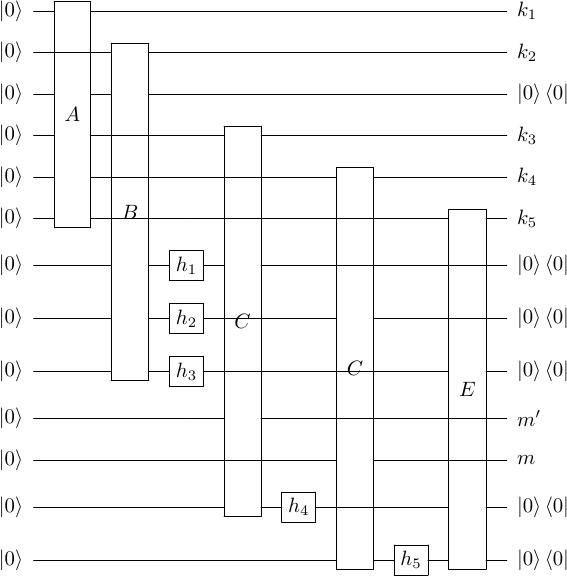}
    % \centering
    % \begin{tikzpicture}[vector/.style={circle, draw},]
    % \draw (90+360/5:2) node[circle, fill] (n1) {} node[shift={(-0.4,0)}] {$\iota_1$};
    % \draw (90+360/5*2:2) node[circle, fill] (n2) {} node[shift={(-0.4,0)}] {$\iota_2$};
    % \draw (90+360/5*3:2) node[circle, fill] (n3) {} node[shift={(0,0.4)}] {$\iota_3$};
    % \draw (90+360/5*4:2) node[circle, fill] (n4) {} node[shift={(0,0.4)}] {$\iota_4$};
    % \draw (90+360/5*5:2) node[circle, fill] (n5) {} node[shift={(-0.4,0)}] {$\iota_5$};
    % \draw[outgoing] (n1) -- (n2) node[pos=0.5, left] {$\mathbb{I}$};
    % \draw[outgoing] (n1) -- (n3) node[pos=0.5, left] {$\mathbb{I}$};
    % \draw[outgoing] (n1) -- (n4) node[pos=0.5, above] {$\mathbb{I}$};
    % \draw[outgoing] (n1) -- (n5) node[pos=0.5, above] {$\mathbb{I}$};
    % \draw[outgoing] (n2) -- (n3) node[pos=0.5, above] {$h_1$};
    % \draw[outgoing] (n2) -- (n4) node[pos=0.5, left] {$h_2$};
    % \draw[outgoing] (n2) -- (n5) node[pos=0.5, left] {$h_3$};
    % \draw[outgoing] (n3) -- (n5) node[pos=0.5, left] {$h_4$};
    % \draw[outgoing] (n4) -- (n5) node[pos=0.5, above] {$h_5$};
    % \draw[outgoing] (n4) -- ++(0.8,0) node[pos=1.3] {$m$};
    % \draw[outgoing] (n3) -- ++(0.8,0) node[pos=1.3] {$m'$};
    % \end{tikzpicture}
    \caption{Spin network corresponding to a pentagram with two open legs.}
    \label{fig:open_pentagram}
\end{figure}

\begin{figure}[ht!]
    \includegraphics[]{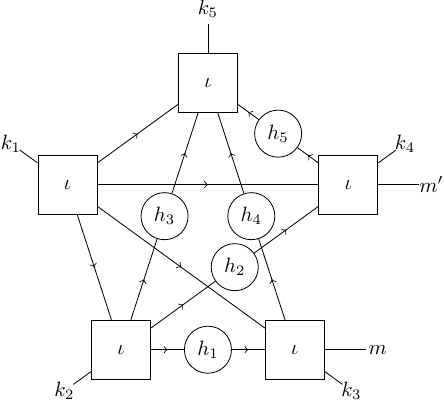}
    % \centering
    % \begin{tikzpicture}[
    % vector/.style={circle, draw},
    % tensor/.style={rectangle, draw, minimum size=10mm},
    % ]
    % \draw (90+360/5:2.5) node[tensor] (n1) {$\iota$};
    % \draw (90+360/5*2:2.5) node[tensor] (n2) {$\iota$};
    % \draw (90+360/5*3:2.5) node[tensor] (n3) {$\iota$};
    % \draw (90+360/5*4:2.5) node[tensor] (n4) {$\iota$};
    % \draw (90+360/5*5:2.5) node[tensor] (n5) {$\iota$};
    % \draw[outgoing] (n1) -- (n2);
    % \draw[outgoing] (n1) -- (n3);
    % \draw[outgoing] (n1) -- (n4);
    % \draw[outgoing] (n1) -- (n5);
    % \draw ($0.5*(n2)+0.5*(n3)$) node[vector] (h1) {$h_1$};
    % \draw[outgoing] (n2) -- (h1);
    % \draw[outgoing] (h1) -- (n3);
    % \draw ($0.5*(n2)+0.5*(n4)$) node[vector] (h2) {$h_2$};
    % \draw[outgoing] (n2) -- (h2);
    % \draw[outgoing] (h2) -- (n4);
    % \draw ($0.5*(n2)+0.5*(n5)$) node[vector] (h3) {$h_3$};
    % \draw[outgoing] (n2) -- (h3);
    % \draw[outgoing] (h3) -- (n5);
    % \draw ($0.5*(n3)+0.5*(n5)$) node[vector] (h5) {$h_4$};
    % \draw[outgoing] (n3) -- (h5);
    % \draw[outgoing] (h5) -- (n5);
    % \draw ($0.5*(n4)+0.5*(n5)$) node[vector] (h6) {$h_5$};
    % \draw[outgoing] (n4) -- (h6);
    % \draw[outgoing] (h6) -- (n5);
    % \draw (n1) -- ++(144:1) node[pos=1.5] {$k_1$};
    % \draw (n2) -- ++(216:1) node[pos=1.5] {$k_2$};
    % \draw (n3) -- ++(-36:1) node[pos=1.5] {$k_3$};
    % \draw (n4) -- ++(36:1) node[pos=1.5] {$k_4$};
    % \draw (n5) -- ++(90:1) node[pos=1.5] {$k_5$};
    % \draw (n3) -- ++(1.2,0) node[pos=1.3] {$m$};
    % \draw (n4) -- ++(1.2,0) node[pos=1.3] {$m'$};
    % \end{tikzpicture}
    \caption{Tensor network corresponding to pentagram spin network 
    with two open legs.}
    \label{fig:open_pentagram_tn}
\end{figure}

It is possible to prepare states of open spin networks 
using the method presented. For this purpose, one can 
leave particular legs of $\iota$ tensors without connecting 
them to other tensors. An example of the case of the 
pentagram spin network is shown in Figs.~\ref{fig:open_pentagram} 
and \ref{fig:open_pentagram_tn}. 

It should be noted that a distinct definition of an open spin 
network is utilized in this context compared to the one presented 
in Ref. \cite{czelusta2023quantum}. There, we considered open links 
as links that are still endowed with holonomy. In other words, 
the link had two spins associated with it, one at the outer node 
and one at the loose end of the link. In the current work, we 
consider an open link as a link without holonomy and only one 
spin, which is part of the outer node. The latter definition 
is common in the literature.

\begin{figure}[ht!]
    \includegraphics[scale=0.8]{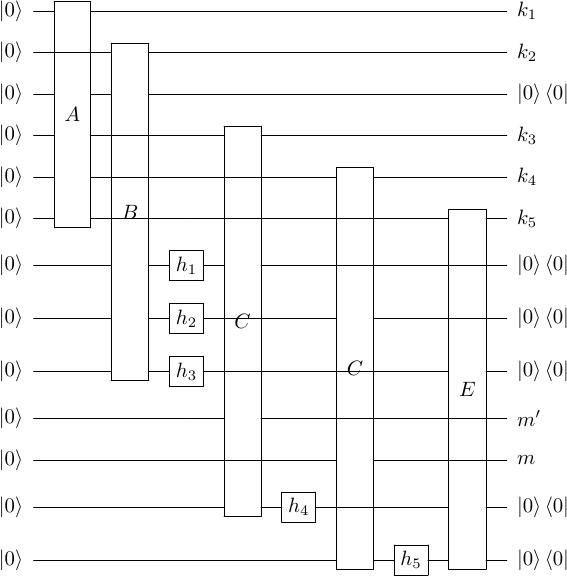}
        \caption{Quantum circuit for the open pentagram spin network
        shown in Fig. \ref{fig:open_pentagram}.}
        \label{fig:openpentagramcirc}
\end{figure}

In Fig. \ref{fig:openpentagramcirc}, the construction of 
the quantum circuit associated with the open spin network 
shown in Fig. \ref{fig:open_pentagram} is presented. As previously, 
an explicit expressions for the multi-qubit gates $A, B, C, D$, 
and $E$ can be found in Appendix A. 

\subsection{Complexity of the construction}

One can transform any graph (without self-loops) into a DAG 
(Direct Acyclic Graph). For this purpose, number the graph nodes 
from $1$ to $n$, and direct each link so that it points from 
a lower-numbered node to a higher-numbered node. Then, by numbering 
the nodes using \emph{topological sorting} \cite{knuth1997art}, one 
can obtain a DAG with a single source (node with only outgoing links) 
and a single sink (node with only ingoing links). The directivity 
of this graph allows it to be represented as a sequence of gates. 
If new edge directions do not match the spin network directions, one 
can simply change the holonomy to its inverse $h\rightarrow h^{-1}$ 
on a given link. Besides source and sink, the remaining nodes are 
of type (1,3), (2,2), or (3,1), where the notation follows 
\textit{(number of incoming links, number of outgoing links)}. One can see 
that the number of nodes $(1,3)$ is the same as the number of
nodes $(3,1)$. Let us denote this number as $x$. Also, let $y$ be 
the number of $(2,2)$ vertices. Then, the following equality holds:
\begin{equation}
1+x+y+x+1=n,\quad 2x+y=n-2.
\end{equation}

One can now analyze the number of qubits required to construct a 
graph with $n$ nodes. Block $A$ needs 5 qubits, and block $B$ needs 
4 new qubits because one is the incoming link, so it is already 
counted in some other block. Block $C$ needs 2 new qubits because 
2 others are already counted, and similarly $D$ needs 1 new qubit 
and $E$ needs no new qubits. Therefore, one needs $5+4x+2y+x=x+2n+1$ 
qubits in total. Because $0\leq x\leq \frac{n-2}{2}$ then number 
of required qubits lies between $2n+1$ and $2n+\frac{n}{2}$.

Such values are explicitly more optimal than the $4n$ number 
of qubits needed to construct arbitrary Ising-type spin network
from considering first all links and then performing the 
projections at the nodes. 

\section{Holographic map}
\label{Sec.BulkBoundary}

The presented technique may prove advantageous for simulating problems
that are challenging for classical numerical methods, as well as for 
improving the representation and comprehension of new concepts in LQG.

One such concept is the \textit{boundary map} \cite{chen2021loop}, 
which is similar to a holographic idea arising from the interpretation 
of open spin networks as a volume (bulk) enclosed by a surface (boundary - $\partial$):
\begin{equation}
\psi : \ket{ \varphi_{\text{bulk}}} \mapsto 
\bra{\psi[\varphi_{\text{bulk}}]} \in \mathcal{H}_{\partial}^{*}, \quad \langle \psi[\varphi_{\text{bulk}}] | \Phi_{\partial} \rangle \in \mathbb{C},
\end{equation}
where $\ket{\Phi_{\partial}} \in \mathcal{H}_{\partial}$ is a state of a boundary. 

The map $\psi$ depends on the bulk state $\ket{\varphi_{\text{bulk}}}$ 
and is valued in the dual of the boundary Hilbert space, $\bra{\psi[\varphi_{\text{bulk}}]} \in \mathcal{H}_{\partial}^{*}$. 

Consequently, it defines a linear form on the boundary Hilbert space. 
Two possible interpretations can be considered. First, the map may 
be interpreted as defining a probability distribution for the bulk
observables, dependent on the choice of boundary states (\emph{i.e.},
quantum boundary conditions). Once $\ket{\Phi_\partial}$ is fixed, 
the function:
\begin{equation}
\langle \Phi_{\partial} | \psi [ \cdot ] \rangle : \ket{\varphi_{\text{bulk}}} \mapsto \langle \Phi_{\partial} | \psi [ \varphi_{\text{bulk}} ] \rangle \in \mathbb{C},
\end{equation}
is the $\mathbb{C}$-valued wave function for the bulk states.

Alternatively, one may reverse this logic and examine the probability 
distribution for the boundary observables after integrating over the 
bulk states. In this case,
\begin{equation}
\hat{\rho}_{\partial}[\psi] = \int \mathcal{D} \varphi_{\text{bulk}} \, |\psi[\varphi_{\text{bulk}}] \rangle \langle \psi[\varphi_{\text{bulk}}]| \in \text{End}(\mathcal{H}_{\partial}),
\end{equation}
represents the density matrix induced by the bulk state $\psi$ 
on the boundary. The measure $\mathcal{D} \varphi_{\text{bulk}}$
can be, in particular, the Haar measure for the qubit systems. 
The averaging of probability distributions over bulk degrees of 
freedom can be achieved by utilizing a quantum computing concept 
known as \textit{unitary design} \cite{gross2007evenly, dankert2009exact}.

Let $P_{t,t}(U)$ be a polynomial with a homogeneous degree at 
most $t$ in $d$ variables in the entries of a unitary matrix $U$ 
and degree $t$ in the complex conjugates of those entries. A 
unitary $t$-design is a set of $K$ unitaries $\{U_k\}$ such that:
\begin{equation}
\frac{1}{K} \sum_{k=1}^{K} P_{t,t}(U_k) = \int_{U(d)} P_{t,t}(U) d\mu(U),
\end{equation}
holds for all possible $P_{t,t}$, and where $d\mu$ is the uniform 
Haar measure.

In \cite{chen2021loop}, the edges of the spin network are interpreted 
as gates (in our construction $h$ gates). The holonomies (group elements) 
along the edges of the graph act as unitary gates in the quantum circuit. 
Each edge represents a transformation associated with a specific $SU(2)$ 
group element. In turn, the nodes correspond to multiqubit gates, which 
are associated with the $A, B, C, D$, and $E$ gates in the representation 
introduced here. Intertwiners are analogous to multi-qubit gates that 
ensure the states respect the local $SU(2)$ gauge invariance. These 
intertwiners connect multiple edges (spins), ensuring the proper entanglement
and correlation among them. This fully aligns with our construction.
A circuit corresponding to the bulk operator can be prepared explicitly 
using our method and executed on a quantum computer.

This method facilitates studying how local geometric information 
propagates through a discrete quantum structure, thus bridging the 
gap between abstract theoretical constructs and practical quantum 
computing techniques. It can support exploring holographic principles 
in LQG, where bulk geometry is encoded on the boundary states, much 
like quantum information is processed in a circuit. It aids in developing
procedures for reconstructing bulk geometry from boundary data, which 
is crucial for understanding how local geometric information propagates 
and correlates. The framework can inspire new quantum computational 
models that simulate aspects of quantum gravity, providing a testing 
ground for theories and computations.

\begin{figure}[ht!]
    \centering
    \begin{subfigure}[b]{0.45\textwidth}
        \includegraphics[]{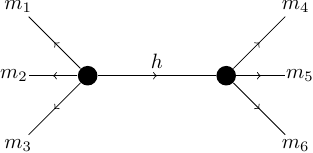}
        % \begin{tikzpicture}
        % \draw node[circle, fill] (n) {};
        % \draw[outgoing] (n) -- ++(-1,1) node[pos=1.2] {$m_1$};
        % \draw[outgoing] (n) -- ++(-1,0) node[pos=1.3] {$m_2$};
        % \draw[outgoing] (n) -- ++(-1,-1) node[pos=1.2] {$m_3$};
        % \draw[outgoing] (n.east) -- ++(2,0) node[pos=0.5, above] {$h$} node[circle, fill, right] (n2) {};
        % \draw[outgoing] (n2) -- ++(1,1) node[pos=1.2] {$m_4$};
        % \draw[outgoing] (n2) -- ++(1,0) node[pos=1.3] {$m_5$};
        % \draw[outgoing] (n2) -- ++(1,-1) node[pos=1.2] {$m_6$};
        % \end{tikzpicture}    
    \end{subfigure}
    \vfill
    \begin{subfigure}[b]{0.45\textwidth}
        \includegraphics[]{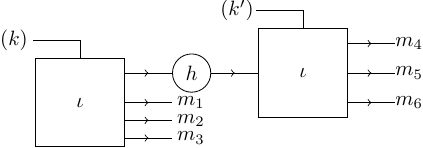}
        % \begin{tikzpicture}[
        % tensor/.style={rectangle, draw, minimum size=15mm},
        % vector/.style={circle, draw},
        % ]
        % \coordinate (a) at (0.8,0);
        % \coordinate (b) at (0.8,0);
        % \node[tensor] (i) {$\iota$};
        % \draw (i.north) -- ++(0,0.3) -- ++(-0.8,0) node[pos=1.4] {$(k)$};
        % \draw[outgoing] ($(i.east)+(0,0)$) -- ++(0.8,0) node[pos=1.4] {$m_1$};
        % \draw[outgoing] ($(i.east)-(0,0.3)$) -- ++(0.8,0) node[pos=1.4] {$m_2$};
        % \draw[outgoing] ($(i.east)-(0,0.6)$) -- ++(0.8,0) node[pos=1.4] {$m_3$};
        % \draw[outgoing] ($(i.east)+(0,0.5)$) -- ++(0.8,0) node[vector, right] (h) {$h$};
        % \draw[outgoing] (h.east) -- ++(0.8,0) node[tensor, right] (i2) {$\iota$};
        % \draw (i2.north) -- ++(0,0.3) -- ++(-0.8,0) node[pos=1.4] {$(k')$};
        % \draw[outgoing] ($(i2.east)+(0,0.5)$) -- ++(0.8,0) node[pos=1.3] {$m_4$};
        % \draw[outgoing] (i2.east) -- ++(0.8,0) node[pos=1.3] {$m_5$};
        % \draw[outgoing] ($(i2.east)+(0,-0.5)$) -- ++(0.8,0) node[pos=1.3] {$m_6$};
        % \end{tikzpicture}
    \end{subfigure}
    \vfill
    \begin{subfigure}[b]{0.45\textwidth}
        \includegraphics[]{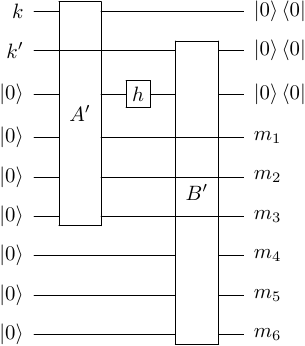}
        % \leavevmode
        % \centering
        % \Qcircuit @C=1.2em @R=1em {
        %     \lstick{k}       & \multigate{5}{A'}  & \qw      & \qw                & \rstick{\ket{0}\bra{0}} \qw \\
        %     \lstick{k'}      & \qw                & \qw      & \multigate{7}{B'}  & \rstick{\ket{0}\bra{0}} \qw \\
        %     \lstick{\ket{0}} & \ghost{A'}         & \gate{h} & \ghost{B'}         & \rstick{\ket{0}\bra{0}} \qw \\
        %     \lstick{\ket{0}} & \ghost{A'}         & \qw      & \qw                & \rstick{m_1} \qw \\
        %     \lstick{\ket{0}} & \ghost{A'}         & \qw      & \qw                & \rstick{m_2} \qw \\
        %     \lstick{\ket{0}} & \ghost{A'}         & \qw      & \qw                & \rstick{m_3} \qw \\
        %     \lstick{\ket{0}} & \qw                & \qw      & \ghost{B'}         & \rstick{m_4} \qw \\
        %     \lstick{\ket{0}} & \qw                & \qw      & \ghost{B'}         & \rstick{m_5} \qw \\
        %     \lstick{\ket{0}} & \qw                & \qw      & \ghost{B'}         & \rstick{m_6} \qw \\
        %     }
    \end{subfigure}

    \caption{Spin network, tensor network, and corresponding quantum circuit for an exemplary graph. Circuits $A'$ and $B'$ are version of $A$ and $B$, respectively, with $k$ as an input. Explicit forms can be found in Appendix A.}
    \label{fig:boundary_map}
\end{figure}

In Fig. \ref{fig:boundary_map} an example of the bulk-boundary map 
is shown. 

The bulk-boundary operator discussed in Ref. \cite{colafranceschi2022holographic} 
is another way of introducing the boundary map.  It is an operator responsible for 
mapping the internal degrees of freedom of nodes ($k$) (bulk) to degrees of freedom 
of open links on the boundary ($m$). Such an operator can be practically realized 
using our scheme. 

\begin{figure}[ht!]
    \includegraphics[]{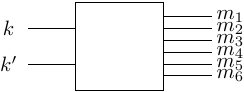}
    % \centering
    % \begin{tikzpicture}[
    % tensor/.style={rectangle, draw, minimum size=15mm},
    % ]
    % \coordinate (a) at (0.8,0);
    % \coordinate (b) at (0.8,0);
    
    % \node[tensor] (i) {};
    % \draw ($(i.west)+(0,0.3)$) -- ++(-0.8,0) node[pos=1.4] {$k$};
    % \draw ($(i.west)-(0,0.3)$) -- ++(-0.8,0) node[pos=1.4] {$k'$};

    % \draw ($(i.east)+(0,0.5)$) -- ++(0.8,0)  node[pos=1.4] {$m_1$};
    % \draw ($(i.east)+(0,0.3)$) -- ++(0.8,0)  node[pos=1.4] {$m_2$};
    % \draw ($(i.east)+(0,0.1)$) -- ++(0.8,0)  node[pos=1.4] {$m_3$};
    % \draw ($(i.east)-(0,0.1)$) -- ++(0.8,0)  node[pos=1.4] {$m_4$};
    % \draw ($(i.east)-(0,0.3)$) -- ++(0.8,0)  node[pos=1.4] {$m_5$};
    % \draw ($(i.east)-(0,0.5)$) -- ++(0.8,0)  node[pos=1.4] {$m_6$};
    % \end{tikzpicture}
    \caption{Diagrammatic representation of the bulk-boundary operator
    associated with the spin network, depicted in Fig. \ref{fig:boundary_map}.}
    \label{fig:bulk_boundary_tensors}
\end{figure}

In Fig. \ref{fig:bulk_boundary_tensors}, an example on the bulk-boundary operator
associated with the pentagram spin network is shown. 

\section{Projection problem}
\label{Sec.ProjectionProblem}

\subsection{Post-selection}
Naive projection on a quantum computer turns out to be inefficient due 
to the need to discard exponentially many measurements. For instance, 
to project onto a state $\ket{0...0}$, all non-$0...0$ measurements 
need rejection. As the number of possible outcomes grows exponentially, 
the probability of attaining the desired state becomes exponentially 
small. Nevertheless, there are potential solutions.

\subsection{Grover's algorithm}

\begin{figure}[ht!]
    \includegraphics[]{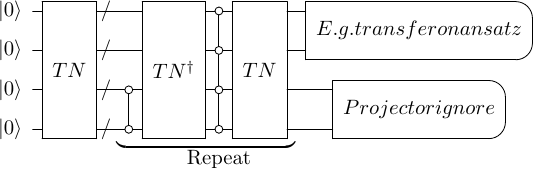}
    % \leavevmode
    % \centering
    % \Qcircuit @C=0.45em @R=1em {
    %     \lstick{\ket{0}}& \multigate{3}{TN} & {/} \qw & \qw & \qw & \multigate{3}{TN^\dagger} & \ctrlo{1}& \multigate{3}{TN} & \qw & \multimeasureD{1}{\text{E.g. transfer on ansatz}}\\
    %     \lstick{\ket{0}}& \ghost{TN} & {/} \qw & \qw & \qw & \ghost{TN^\dagger} & \ctrlo{1}& \ghost{TN} & \qw & \ghost{\text{E.g. transfer on ansatz}}\\
    %     \lstick{\ket{0}}& \ghost{TN} & {/} \qw & \qw & \ctrlo{1} & \ghost{TN^\dagger} & \ctrlo{1}& \ghost{TN} & \qw & \multimeasureD{1}
    %     {\text{Project or ignore}}\\
    %     \lstick{\ket{0}}& \ghost{TN} & {/} \qw & \qw & \controlo \qw & \ghost{TN^\dagger} & \controlo \qw& \ghost{TN} & \qw & \ghost{\text{Project or ignore}}\\
    %      &&&&&&\mbox{Repeat}&\\
    %     {\gategroup{1}{5}{4}{8}{.8em}{_)}}
    %     }
    \caption{Grover's algorithm applied to the projection task.}
    \label{fig:grover}
\end{figure}

Grover's algorithm could be used to amplify states that need to be projected.  
This can reduce the number of rejected measurements, and in an ideal situation, 
we may not need to project at all. Instead, we could ignore those qubits because 
they would all be in the desired state. The general scheme of this method is 
illustrated in Fig. \ref{fig:grover}.

The $TN$ gate represents the complete circuit used to generate a given spin network, 
such as the pentagram shown in Fig. \ref{fig:pentagramcirc}. This circuit operates 
on four qubits in the presented scheme, but it is adaptable to however many qubits 
are required (12 for the pentagram, for instance). Qubits on the scheme are divided 
into two groups: the first two correspond to qubits that carry the obtained spin 
network state (for a pentagram, these are 5 qubits $k$), and the last two correspond 
to projecting qubits (for a pentagram, 7 qubits). Following the usual Grover construction, 
we first mark the desired state by applying an anticontrol-$Z$ gate on the projecting 
qubits, which gives a minus sign to states that have $\ket{0...0}$. We then apply 
reflection to amplify these states. The reflection is obtained using the usual method. 
We apply the $TN^\dagger$, then the anticontrol-$Z$, and then the $TN$ to all qubits. 
The entire Grover algorithm can be repeated until we reach a state close enough to 
the desired state.

This approach can significantly decrease the required number of measurements, 
potentially eliminating the need for them altogether. However, the drawback is the 
demand for a more complex circuit, which may be unattainable with noisy quantum 
processors.

\subsection{Global and local cost functions}

The transfer of the obtained state to a parametrized ansatz, 
the contraction of the network, using variational methods, 
can be obtained by minimizing some cost function. We can use 
a global cost function, which involves the projection of the 
whole state using all qubits:
\begin{equation}
    C_G(\theta)=1-\langle\psi|0...0\rangle.
\end{equation}

The global cost function can be difficult to compute because 
its value is usually exponentially small as the number of 
qubits increases. Therefore, one can consider a local cost function:
\begin{equation}
    C_L(\theta)=1-\sum_n\langle\psi|0_n\rangle,
\end{equation}
which involves projections of each qubit separately.

This is easier to compute, and because of the 
following inequality: 
\begin{equation}
    0\leq C_G(\theta)\leq nC_L(\theta),
\end{equation}
minimizing the local cost function implies minimizing 
the global cost function.

\subsection{Two-way quantum computing}

Recently, a new speculative approach called two-way 
quantum computing (2WQC) has been proposed in Ref. 
\cite{duda2023two}. The core assumption of 2WQC is 
that a desired postselection of the final states can 
be imposed. While the physical feasibility of such an 
operation is under debate, it has already been 
shown that the 2WQC might have profound implications 
on computational complexity. In particular,
within 2WQC, the Grover’s algorithm has a constant 
complexity \cite{czelusta2024grover}. In our context, it 
implies that the projection can performed in just 
a single shot. 

\section{Applications}
\label{Sec.Applications}

\subsection{Pentagram}

Here, we provide an exemplary application of the 
method developed by studying the pentagram spin network. 
Here, the quantum circuit corresponding to the spin network 
is the one shown in Fig. \ref{fig:pentagramcirc}. Following 
the transfer method discussed in Ref. \cite{khatri2019quantum}
and applied in our earlier article \cite{czelusta2023quantum},
we transfer the 5-qubit state of intertwines onto the ansatz
circuit built from the three layers of the circuit shown in Fig. 
\ref{fig:s2d}. The ansatz is the so-called SimplifiedTwoDesign 
\cite{cerezo2021cost}.

The optimization method applied in the procedure involves 
the cost functions:
\begin{equation}
\text{Cost}\left(\vec{\theta}\right) = 1-\text{Prob}(00000),
\end{equation}
where the parameter vector $\vec{\theta}=(\theta_0,\dots,\theta_{28})$.
The initial layer contains five parameters, and each of the three 
layers contains a set of eight parameters. Therefore, the total 
number of parameters in the ansatz is $29$.

\begin{figure}
    \centering
    \includegraphics[scale=0.4]{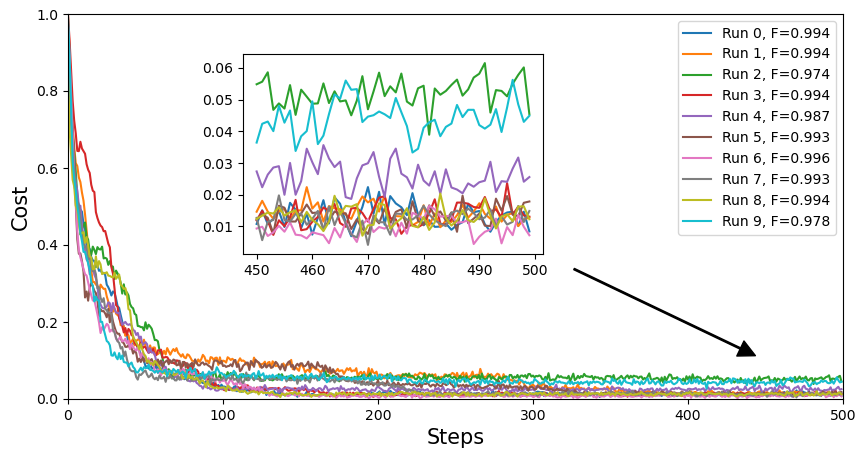}
    \caption{History of the cost function for 10 runs with random initial parameters.}
    \label{fig:cost_pentagra}
\end{figure}

In the classical optimization part, the Adam \cite{kingma2014adam}
minimizing algorithm has been applied. Furthermore, the parameter 
shift rule \cite{mitarai2018quantum, schuld2019evaluating} has been 
used in the computations. In Fig. \ref{fig:cost_pentagra}, the 
time evolution of the cost function, from ten randomly chosen initial 
sets of parameters $(\theta_0,\dots,\theta_{28})$ is shown. 

In each run, the number of iteration steps was 500.
The 12-qubit circuit, at each step of the iteration procedure, 
was executed 20000 times to find the value of the cost function. 
This does not include the additional measurements made for 
the purpose of the gradient finding with the use of the parameter 
shift rule.

Quantum fidelity of the final state of the pentagram, 
averaged over the ten rounds, is $F=0.9897 \pm 0.0073$.

The codes and data that reproduce the results presented 
here are available in the GitHub repository \cite{czelusta2024github}.

\subsection{Holographic bulk-boundary map}

As another example of the application of the constructed 
tensor networks, we present computations of the mutual 
information between two regions of boundary induced by 
a bulk-boundary map. For this purpose, let us consider the 
map from Fig. \ref{fig:boundary_map}. One can compute mutual 
information between boundary degrees of freedom $m_1,m_2,m_3$ 
and $m_4,m_5,m_6$ and find probability distribution over 
these values for random bulk states.

In what follows, we will refer to the entanglement entropy:
\begin{equation}
    S(\hat{\rho}) := - \text{tr}(\hat{\rho}\log_2 \hat{\rho}),
\end{equation}
where the base of the logarithm has been chosen to $2$,
for simplicity of interpretation (which is different from 
the $e$ factor typically used in the definition). 
Here $\hat{\rho}$ is the density matrix of the state under 
consideration. 

For a system $N$-qubit system, subdivided into $A$ and $B$
parts with $N_A$ and $N_B$ qubits correspondingly, the 
Hilbert space can be written as a tensor product:
\begin{equation}
    \mathcal{H}=\mathcal{H}_A\otimes\mathcal{H}_B.
\end{equation}
By tracing over complementary parts, the reduced 
density matrices can be introduced: 
\begin{align}
\hat{\rho}_A := \text{tr}_B(\hat{\rho}), \ \   
\hat{\rho}_B := \text{tr}_A(\hat{\rho}),
\end{align}
so the maximal entropy of subsystem $A$ is equal
$\max S(\hat{\rho}_A) = N_A\log_2 2 = N_A$ and analogously for 
subsystem $B$.

The amount of information present in the entanglement between 
the subsystems $A$ and $B$ can be quantified via the mutual 
information:
\begin{equation}
    I(\hat{\rho}_A,\hat{\rho}_B) := S(\hat{\rho}_A)+S(\hat{\rho}_B)-S(\hat{\rho}).
\end{equation}

\begin{figure}[ht!]
    \includegraphics[]{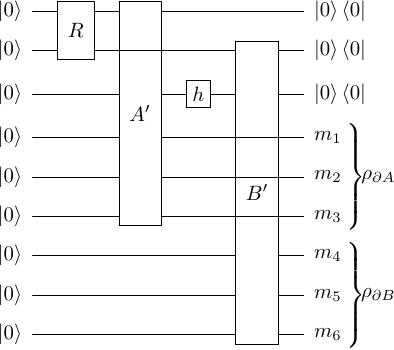}
    % \leavevmode
    % \centering
    % \Qcircuit @C=1.2em @R=1em {
    %     \lstick{\ket{0}} & \multigate{1}{R}                & \multigate{5}{A'}  & \qw      & \qw                & \rstick{\ket{0}\bra{0}} \qw \\
    %     \lstick{\ket{0}} & \ghost{R}                & \qw                & \qw      & \multigate{7}{B'}  & \rstick{\ket{0}\bra{0}} \qw \\
    %     \lstick{\ket{0}} & \qw                & \ghost{A'}         & \gate{h} & \ghost{B'}         & \rstick{\ket{0}\bra{0}} \qw \\
    %     \lstick{\ket{0}} & \qw                & \ghost{A'}         & \qw      & \qw                & \rstick{m_1} \qw \\
    %     \lstick{\ket{0}} & \qw                & \ghost{A'}         & \qw      & \qw                & \rstick{m_2} \qw &&&\mbox{$\rho_{\partial A}$}\\
    %     \lstick{\ket{0}} & \qw                & \ghost{A'}         & \qw      & \qw                & \rstick{m_3} \qw \\
    %     \lstick{\ket{0}} & \qw                & \qw                & \qw      & \ghost{B'}         & \rstick{m_4} \qw \\
    %     \lstick{\ket{0}} & \qw                & \qw                & \qw      & \ghost{B'}         & \rstick{m_5} \qw &&&\mbox{$\rho_{\partial B}$}\\
    %     \lstick{\ket{0}} & \qw                & \qw                & \qw      & \ghost{B'}         & \rstick{m_6} \qw \gategroup{5}{6}{5}{6}{5em}{\}}  \gategroup{8}{6}{8}{6}{5em}{\}}\\
    %     }
        \caption{Procedure of obtaining boundary state from random bulk state. Circuits $A'$ and $B'$ are version of $A$ and $B$, respectively, with $k$ as an input. Explicit forms can be found in Appendix A.}
        \label{fig:bulk_boundary_example}
\end{figure}

The task is now to compute the mutual information for the 
boundary map represented by the circuit shown in Fig. 
\ref{fig:boundary_map}. For this purpose, we randomly 
select the 2-qubit bulk state according to the Haar 
measure. This can be performed by inserting, as the $kk'$, 
a circuit $R$ generating random states.

\begin{figure}[ht!]
    \includegraphics[scale=0.8]{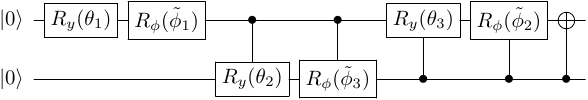}
    % \leavevmode
    % \centering
    % \Qcircuit @C=0.5em @R=1em {
    %     \lstick{\ket{0}} & \gate{R_y(\theta_1)} &\gate{R_\phi(\tilde{\phi}_1)} &\ctrl{1}             &\ctrl{1}                         &\gate{R_y(\theta_3)} &\gate{R_\phi(\tilde{\phi}_2)} &\targ     &\qw\\
    %     \lstick{\ket{0}} & \qw                  &\qw                   &\gate{R_y(\theta_2)} &\gate{R_\phi(\tilde{\phi}_3)} &\ctrl{-1}                 &\ctrl{-1}                        &\ctrl{-1} &\qw\\  
    %     }
    \caption{Quantum circuit used for generating random 2-qubit states. Here: $\tilde{\phi_1}=\phi_1$, $\tilde{\phi_2}=\phi_2-\phi_3+\pi$, and $\tilde{\phi_3}=\phi_3-\phi_1-\pi$.}
    \label{fig:rand_circ}
\end{figure}

An arbitrary two-qubit state can be parametrized using six parameters:
\begin{align}
    \ket{\psi}=&\cos{\left(\frac{\theta_{1}}{2} \right)}\ket{00}+\\
                &e^{i \phi_{1}} \sin{\left(\frac{\theta_{1}}{2} \right)} \cos{\left(\frac{\theta_{2}}{2} \right)}\ket{01}+\\
                &e^{i \phi_{3}} \sin{\left(\frac{\theta_{1}}{2} \right)} \sin{\left(\frac{\theta_{2}}{2} \right)} \cos{\left(\frac{\theta_{3}}{2} \right)}\ket{10}+\\
                &e^{i \phi_{3}} \sin{\left(\frac{\theta_{1}}{2} \right)} \sin{\left(\frac{\theta_{2}}{2} \right)} \sin{\left(\frac{\theta_{3}}{2} \right)}\ket{11},\\
\end{align}
where $\theta_k\in [0,\pi],\; \phi_k\in[0,2\pi)$, which can be generated 
using the circuit shown in Fig. \ref{fig:rand_circ}.

To sample states according to Haar measure we need to sample parameters $\theta$ and $\phi$ according to proper measure \cite{zyczkowski2001induced}.
Parameters $\theta_k$ must be sampled from:
\begin{equation}
    P(\theta_k)=k\sin\theta_k\left(\sin\frac{\theta_k}{2}\right)^{2k-2},
\end{equation}
and parameters $\phi_k$ must be sampled from uniform distribution:
\begin{equation}
    P(\phi_k)=\frac{1}{2\pi}.
\end{equation}

Using PennyLane \cite{bergholm2018pennylane}, we prepare a circuit $R$ from Fig. \ref{fig:rand_circ}:

\begin{lstlisting}[language=Python]
import pennylane as qml

def rand_state(thetas, phis, wires):
    qml.RY(thetas[0],wires[0])
    qml.PhaseShift(phis[0],wires[0])
    qml.CRY(thetas[1], wires)
    qml.CPhase(phis[2]-phis[0]-np.pi, wires)
    qml.CRY(thetas[2], wires[::-1])
    qml.CPhase(phis[1]-phis[2]+np.pi, wires[::-1])
    qml.CNOT(wires[::-1])
\end{lstlisting}
and circuit performing the whole procedure from
Fig. \ref{fig:bulk_boundary_example}:

\begin{lstlisting}[language=Python]
dev = qml.device('default.qubit', wires = 9)

@qml.qnode(dev)
def circ(thetas, phis):
    rand_state(thetas, phis, wires=[0,1])
    circAA(wires=[0,2,3,4,5])
    circBB(wires=[1,2,6,7,8])
    qml.measure(0, postselect=0)
    qml.measure(1, postselect=0)
    qml.measure(2, postselect=0)
    return qml.mutual_info(wires0=[3,4,5], wires1=[6,7,8], log_base=2)
\end{lstlisting}

Generating boundary states for different parameters $\theta_k$ and $\phi_k$ we obtain distribution of mutual information:

\begin{figure}[ht!]
    \centering
    \includegraphics[scale=0.4]{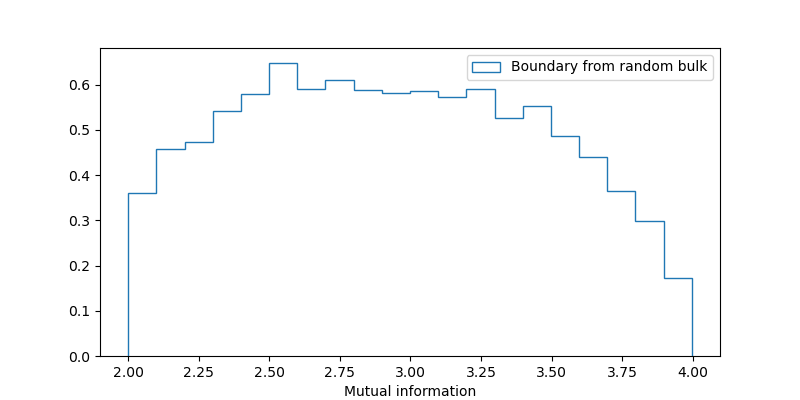}
    \caption{Distribution of mutual information between boundary degrees of freedom $m_1,m_2,m_3$ and $m_4,m_5,m_6$ for random bulk state.}
    \label{fig:boundary_hist}
\end{figure}

We can plot these results together with the distribution 
of mutual information for bulk degrees of freedom and for 
random boundary state (not obtained from random bulk):
\begin{figure}[ht!]
    \centering
    \includegraphics[scale=0.4]{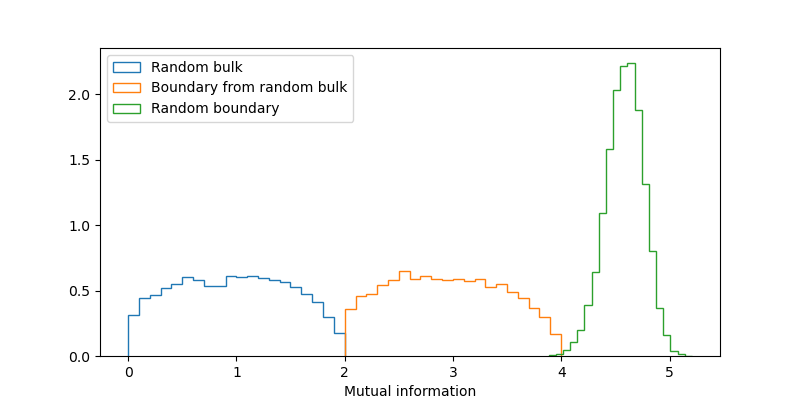}
    \caption{Distributions of mutual information for: random bulk state, boundary state computed from the 
    random bulk state, and for random boundary state.}
    \label{fig:enter-label}
\end{figure}

As we can see, the distribution for the boundary obtained by 
the bulk-boundary map is completely different from that for 
the random boundary and has the same shape as the distribution 
for the bulk. It is just shifted by 2 to the higher values of 
mutual information. Furthermore, the distribution is independent 
of holonomy $h$.

Our result is consistent with equation (17) from Ref. \cite{livine2018intertwiner}, 
which states that the entanglement entropy of boundary spins differs 
from the entanglement entropy of bulk nodes by a constant depending 
only on spin $j$ of the link between bulk nodes:
\begin{equation}
    S(\hat{\rho}_{\partial A}) = S(\hat{\rho}_A) + \log\left(2j+1\right).
\end{equation}

In our case $j=1/2$ and we use base two logarithms, so:
\begin{equation}
    S(\hat{\rho}_{\partial A}) = S(\hat{\rho}_A) + 1,
\end{equation}
and consequently, the mutual information satisfies:
\begin{equation}
    I(\hat{\rho}_{\partial A},\hat{\rho}_{\partial B}) = I(\hat{\rho}_A,\hat{\rho}_B) + 2,
\end{equation}
which is exactly why the distribution of mutual information 
for the boundary is shifted by the factor ``2" relative to the 
distribution for bulk.

A detailed discussion of the entanglement entropy, employing the 
tensor network methods will be discussed elsewhere. 

The codes and data that can be used to reproduce the results 
presented here are available in the GitHub repository 
\cite{czelusta2024github}.

\section{Summary}
\label{Sec.Summary}

The article is our next step in the research program, aiming at 
performing large-scale quantum simulations of the models of 
LQG. Such simulations, if eventually executed, would allow us 
to study the collective properties of the basic -- Planck scale -- 
degrees of freedom. This, in particular, includes aspects such
as the thermodynamic limit, semiclassical limit, and the 
correspondence to GR, as well as the possible phases
of the gravitational field and the emergence of the holographic 
principle. An important question also corresponds to the local 
geometric fluctuations, which may lead to the formation of structures 
in the Universe (see, e.g., Ref. \cite{Gozzini:2019nbo} for the 
first attempts in this direction). 

Approaching these problems via quantum simulations requires 
extensive quantum computing resources that are not yet available. 
While the quantum circuit representation of the Ising spin network 
states has already been introduced in Refs. \cite{Czelusta:2020ryq,czelusta2023quantum},
the construction is not necessarily an optimal one. Hence, searching 
for implementations involving fewer quantum resources opens a perspective 
for faster achievement of intended research goals. 

In this article, we introduced a tensor network representation 
of the Ising spin networks, which involved fewer qubits in constructing 
quantum circuits than the previous methods. Therefore, the method 
prognoses as a tool to perform cutting-edge quantum simulations 
of the spin networks. As a glimpse of possible applications of the 
method, computations of the entanglement entropy for holographic
configurations of the open spin network are made. The computations 
also provide a cross-check for the methods since the results have 
been computed using other approaches. 

Future work in this area must include the implementation of the Hamiltonian 
constraint to extract the physical states of the theory. Progress in this 
direction has already been initiated \cite{Czelusta:2021jro, grabowska2024fully}. 
Another significant path of research involves exploring the complexity 
of the states under consideration, which is pivotal for emulating quantum 
systems on classical computers. Recent advancements in this domain 
\cite{Cepollaro:2024qln} have examined the complexity of quantum geometry 
through the stabilizer entropy of $SU(2)$ intertwiners, shedding light 
on the computational challenges associated with these states.

A certain drawback of the tensor network-based method of constructing 
quantum circuits is the need for projections. The projections require 
numerous executions of the quantum circuit to obtain the desired 
state on which the projection is made.  The projected state is on the 
subspace of the whole Hilbert space under consideration. The article 
has studied a few approaches to mitigate the problem. Specifically, 
we have shown that when universal quantum computing resources are
available, the Grover algorithm can speed up searching for the 
projected state. Furthermore, the variational method involving 
the local cost functions may provide improvement. Finally, we 
also invoked the recently introduced idea of two-way quantum 
computing (2WQC), which could also improve the method. However, 
the approach is still speculative, and its practical applicability 
remains unclear.

Furthermore, the introduced tensor network representation is 
discussed in the context of holographic bulk-boundary correspondence. 
It is shown that the tensor network-based construction provides 
a natural framework for introducing such concepts as the 
boundary map and the bulk-boundary operator. The considerations 
are supported with an example for which the entanglement entropy
is studied. The analysis provides a numerical confirmation of 
the formula relating bulk and boundary entropies for open spin
networks, derived in Ref. \cite{livine2018intertwiner}. 
More general analysis of the subject, beyond the example 
under consideration, will be made in our forthcoming
studies. 

Finally, let us emphasize that while the considerations presented 
here are solely reserved for the $j=1/2$ case and the four valent 
nodes, extending the method to the higher representations 
is natural in the tensor network approach. This is especially 
interesting from the perspective of the semi-classical limit, which 
is expected to be approached for $j \rightarrow \infty$. 

\section*{Acknowledgements}
This research was funded by the National Science Centre, 
Poland, grant number 2023/49/N/ST2/03845.

\section*{Appendix A}
\label{sec:appendix}

In the following circuits gates containing only number $\theta$ are $R_y(\theta)$ 
gates, with $\alpha=2\arccos\frac{1}{\sqrt{3}}$. Furthermore, the symbols commonly 
used to depict other types of gates are used (e.g., CNOT, $X$, $Z$, $H$, $S$, and $T$).   

\begin{figure}[ht!]
    \includegraphics[]{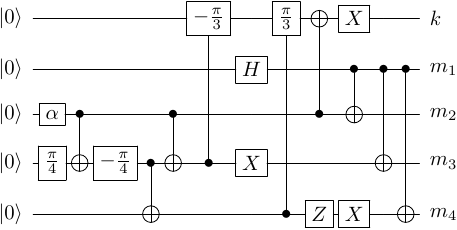}
    % \leavevmode
    % \centering
    % \Qcircuit @C=0.25em @R=1em {
    %     \lstick{\ket{0}}&  \qw & \qw & \qw &  \qw &  \qw &\gate{-\frac{\pi}{3}} &  \qw &\gate{\frac{\pi}{3}}&\targ&\gate{X}&  \qw &  \qw &\rstick{k} \qw \\
    %     \lstick{\ket{0}}&  \qw & \qw & \qw &  \qw &  \qw &\qw        &  \gate{H} &  \qw &  \qw &\ctrl{1}&\ctrl{2}&\ctrl{3}&\rstick{m_1} \qw\\
    %     \lstick{\ket{0}}&  \gate{\alpha} & \ctrl{1}& \qw & \qw &\ctrl{1}&  \qw &  \qw &  \qw &\ctrl{-2}&\targ&  \qw &  \qw & \rstick{m_2} \qw\\
    %     \lstick{\ket{0}}&  \gate{\frac{\pi}{4}} &\targ &  \gate{-\frac{\pi}{4}} & \ctrl{1} &\targ &\ctrl{-3}&\gate{X}&  \qw &  \qw &   \qw &\targ&  \qw &\rstick{m_3} \qw\\
    %     \lstick{\ket{0}}&  \qw & \qw & \qw & \targ &  \qw &  \qw &  \qw &\ctrl{-4}&\gate{Z}&\gate{X}&  \qw &\targ& \rstick{m_4} \qw
    %     }
    \caption{Quantum circuit $A$ corresponding to the tensor $\iota_{(k)}^{m_1m_2m_3m_4}$.}
    \label{fig:A_explicit}
\end{figure}

\begin{figure}[ht!]
    \includegraphics[scale=0.9]{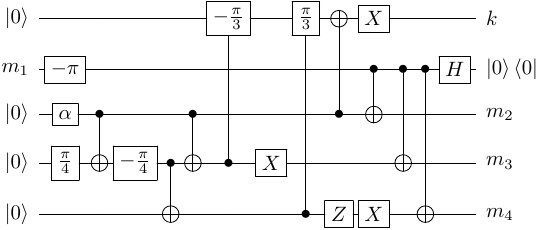}
    % \leavevmode
    % \centering
    % \Qcircuit @C=0.25em @R=1em {
    %     \lstick{\ket{0}}&  \qw & \qw & \qw &  \qw &  \qw&\gate{-\frac{\pi}{3}} &  \qw &\gate{\frac{\pi}{3}}&\targ&\gate{X}&  \qw &  \qw &  \qw &\rstick{k} \qw \\
    %     \lstick{m_1}&  \gate{-\pi} & \qw &\qw &  \qw &  \qw &  \qw &  \qw &  \qw &  \qw &\ctrl{1}&\ctrl{2}&\ctrl{3}&\gate{H}&\rstick{\ket{0}\bra{0}} \qw\\
    %     \lstick{\ket{0}}&  \gate{\alpha} & \ctrl{1}& \qw & \qw &\ctrl{1}&  \qw &  \qw &  \qw &\ctrl{-2}&  \targ&\qw &  \qw &  \qw & \rstick{m_2} \qw\\
    %     \lstick{\ket{0}}&  \gate{\frac{\pi}{4}} &\targ &  \gate{-\frac{\pi}{4}} & \ctrl{1} &\targ &\ctrl{-3}&\gate{X}&  \qw &  \qw &  \qw &  \targ&\qw &  \qw &\rstick{m_3} \qw\\
    %     \lstick{\ket{0}}&  \qw & \qw & \qw & \targ &  \qw &  \qw &  \qw &\ctrl{-4}&\gate{Z}&\gate{X}&  \qw &  \targ&\qw & \rstick{m_4} \qw
    %     }
    \caption{Quantum circuit $B$ corresponding to the tensor 
 $\iota\indices{_{(k)m_1}^{m_2m_3m_4}}$.}
    \label{fig:B_explicit}
\end{figure}

\begin{figure}[ht!]
    \includegraphics[scale=0.9]{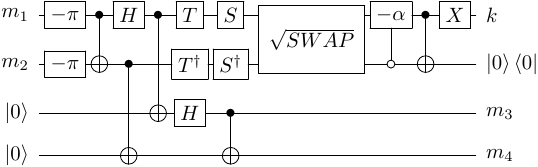}
    % \leavevmode
    % \centering
    % \Qcircuit @C=0.25em @R=1em {
    %     \lstick{m_1}& \gate{-\pi}& \ctrl{1}& \gate{H} & \ctrl{2}& \gate{T}& \gate{S}&  \qw & \multigate{1}{\sqrt{SWAP}}& \gate{-\alpha}& \ctrl{1} &\gate{X}&\rstick{k}\qw\\
    %     \lstick{m_2}& \gate{-\pi}& \targ& \ctrl{2} & \qw& \gate{T^\dagger}& \gate{S^\dagger}& \qw & \ghost{\sqrt{SWAP}} & \ctrlo{-1} & \targ & \qw&\rstick{\ket{0}\bra{0}}\qw\\
    %     \lstick{\ket{0}} & \qw&\qw& \qw& \targ& \gate{H} &  \ctrl{1} & \qw & \qw & \qw & \qw & \qw&\rstick{m_3}\qw\\
    %     \lstick{\ket{0}} & \qw&\qw& \targ& \qw & \qw & \targ& \qw & \qw & \qw & \qw & \qw&\rstick{m_4}\qw\\
    % }
    \caption{Quantum circuit $C$ corresponding to the tensor 
 $\iota\indices{_{(k)}_{m_1m_2}^{m_3m_4}}$.}
    \label{fig:C_explicit}
\end{figure}

\begin{figure}[ht!]
    \includegraphics[scale=0.9]{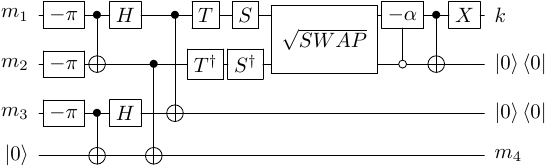}
    % \leavevmode
    % \centering
    % \Qcircuit @C=0.2em @R=1em {
    %     \lstick{m_1}& \gate{-\pi}& \ctrl{1}& \gate{H} & \qw & \ctrl{2}& \gate{T}& \gate{S}&  \qw & \multigate{1}{\sqrt{SWAP}}& \gate{-\alpha}& \ctrl{1} &\gate{X}&\rstick{k} \qw\\
    %     \lstick{m_2}&\gate{-\pi}& \targ& \qw & \ctrl{2} & \qw& \gate{T^\dagger}& \gate{S^\dagger}& \qw & \ghost{\sqrt{SWAP}} & \ctrlo{-1} & \targ & \qw&\rstick{\ket{0}\bra{0}}\qw\\
    %     \lstick{m_3} & \gate{-\pi}& \ctrl{1}& \gate{H}& \qw& \targ& \qw & \qw & \qw & \qw & \qw & \qw & \qw&\rstick{\ket{0}\bra{0}}\qw\\
    %     \lstick{\ket{0}} &\qw & \targ&\qw& \targ& \qw & \qw & \qw & \qw & \qw & \qw & \qw & \qw&\rstick{m_4}\qw\\
    % }
    \caption{Quantum circuit $D$ corresponding to the tensor 
 $\iota\indices{_{(k)}_{m_1m_2m_3}^{m_4}}$.}
    \label{fig:D_explicit}
\end{figure}

\begin{figure}[ht!]
    \includegraphics[scale=0.9]{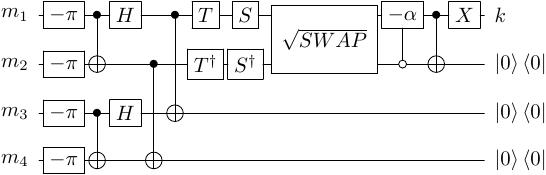}
    % \leavevmode
    % \centering
    % \Qcircuit @C=0.2em @R=1em {
    %     \lstick{m_1}& \gate{-\pi}& \ctrl{1}& \gate{H} & \qw & \ctrl{2}& \gate{T}& \gate{S}&  \qw & \multigate{1}{\sqrt{SWAP}}& \gate{-\alpha}& \ctrl{1} &\gate{X}&\rstick{k} \qw\\
    %     \lstick{m_2}&\gate{-\pi}& \targ& \qw & \ctrl{2} & \qw& \gate{T^\dagger}& \gate{S^\dagger}& \qw & \ghost{\sqrt{SWAP}} & \ctrlo{-1} & \targ & \qw&\rstick{\ket{0}\bra{0}}\qw\\
    %     \lstick{m_3} & \gate{-\pi}& \ctrl{1}& \gate{H}& \qw& \targ& \qw & \qw & \qw & \qw & \qw & \qw & \qw&\rstick{\ket{0}\bra{0}}\qw\\
    %     \lstick{m_4} &\gate{-\pi} & \targ&\qw& \targ& \qw & \qw & \qw & \qw & \qw & \qw & \qw & \qw&\rstick{\ket{0}\bra{0}}\qw\\
    % }
    \caption{Quantum circuit $E$ corresponding to the tensor 
 $\iota\indices{_{(k)}_{m_1m_2m_3m_4}}$.}
    \label{fig:E_explicit}
\end{figure}

\begin{figure}[ht!]
    \includegraphics[scale=0.9]{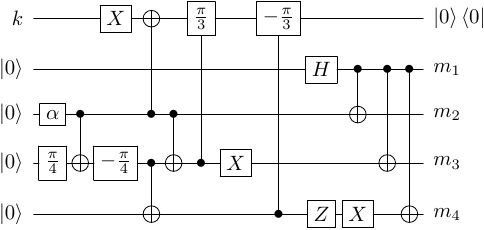}
    % \leavevmode
    % \centering
    % \Qcircuit @C=0.25em @R=1em {
    %     \lstick{k}&  \qw & \qw & \gate{X} &  \targ &  \qw&\gate{\frac{\pi}{3}} &  \qw &\gate{-\frac{\pi}{3}}&\qw&\qw&  \qw &  \qw &\rstick{\ket{0}\bra{0}} \qw \\
    %     \lstick{\ket{0}}&  \qw & \qw &\qw &  \qw &  \qw &  \qw &  \qw &  \qw &\gate{H}&\ctrl{1}&\ctrl{2}&\ctrl{3}&\rstick{m_1} \qw\\
    %     \lstick{\ket{0}}&  \gate{\alpha} & \ctrl{1}& \qw & \ctrl{-2} &\ctrl{1}&  \qw &  \qw &  \qw &  \qw &\targ&  \qw &  \qw & \rstick{m_2} \qw\\
    %     \lstick{\ket{0}}&  \gate{\frac{\pi}{4}} &\targ &  \gate{-\frac{\pi}{4}} & \ctrl{1} &\targ &\ctrl{-3}&\gate{X}&  \qw  &  \qw &  \qw &\targ&  \qw &\rstick{m_3} \qw\\
    %     \lstick{\ket{0}}&  \qw & \qw & \qw & \targ &  \qw &  \qw &  \qw &\ctrl{-4}&\gate{Z}&\gate{X}&  \qw &\targ& \rstick{m_4} \qw
    %     }
    \caption{Quantum circuit $A'$ corresponding to the tensor $\iota_{(k)}^{m_1m_2m_3m_4}$ but with $k$ as an input.}
    \label{fig:Ap_explicit}
\end{figure}

\begin{figure}[ht!]
    \includegraphics[scale=0.9]{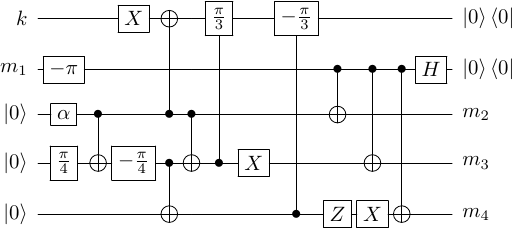}
    % \leavevmode
    % \centering
    % \Qcircuit @C=0.25em @R=1em {
    %     \lstick{k}&  \qw & \qw & \gate{X} & \targ &  \qw&\gate{\frac{\pi}{3}} &  \qw &\gate{-\frac{\pi}{3}}&\qw&\qw&  \qw &  \qw &\rstick{\ket{0}\bra{0}} \qw \\
    %     \lstick{m_1}&  \gate{-\pi} & \qw &\qw &  \qw &  \qw &  \qw &  \qw &  \qw  &\ctrl{1}&\ctrl{2}&\ctrl{3}&\gate{H}&\rstick{\ket{0}\bra{0}} \qw\\
    %     \lstick{\ket{0}}&  \gate{\alpha} & \ctrl{1}& \qw & \ctrl{-2} &\ctrl{1}&  \qw &  \qw &  \qw & \targ&\qw &  \qw &  \qw & \rstick{m_2} \qw\\
    %     \lstick{\ket{0}}&  \gate{\frac{\pi}{4}} &\targ &  \gate{-\frac{\pi}{4}} & \ctrl{1} &\targ &\ctrl{-3}&\gate{X}&  \qw &   \qw &  \targ&\qw &  \qw &\rstick{m_3} \qw\\
    %     \lstick{\ket{0}}&  \qw & \qw & \qw & \targ &  \qw &  \qw &  \qw &\ctrl{-4}&\gate{Z}&\gate{X}&  \targ&\qw & \rstick{m_4} \qw
    %     }
    \caption{Quantum circuit $B'$ corresponding to the tensor $\iota\indices{_{(k)m_1}^{m_2m_3m_4}}$ but with $k$ as an input.}
    \label{fig:Bp_explicit}
\end{figure}

\bibliography{bibliography}

\begin{thebibliography}{42}
\providecommand{\natexlab}[1]{#1}
\providecommand{\url}[1]{\texttt{#1}}
\expandafter\ifx\csname urlstyle\endcsname\relax
  \providecommand{\doi}[1]{doi: #1}\else
  \providecommand{\doi}{doi: \begingroup \urlstyle{rm}\Url}\fi

\bibitem[Orus(2014)]{Orus:2013kga}
Roman Orus.
\newblock {A Practical Introduction to Tensor Networks: Matrix Product States
  and Projected Entangled Pair States}.
\newblock \emph{Annals Phys.}, 349:\penalty0 117--158, 2014.
\newblock \doi{10.1016/j.aop.2014.06.013}.

\bibitem[Biamonte(2019)]{Biamonte:2019uzx}
Jacob Biamonte.
\newblock {Lectures on Quantum Tensor Networks}, 12 2019.

\bibitem[Vidal(2007)]{Vidal:2007hda}
G.~Vidal.
\newblock {Entanglement Renormalization}.
\newblock \emph{Phys. Rev. Lett.}, 99\penalty0 (22):\penalty0 220405, 2007.
\newblock \doi{10.1103/PhysRevLett.99.220405}.

\bibitem[Vidal(2008)]{Vidal:2008zz}
G.~Vidal.
\newblock {Class of Quantum Many-Body States That Can Be Efficiently
  Simulated}.
\newblock \emph{Phys. Rev. Lett.}, 101:\penalty0 110501, 2008.
\newblock \doi{10.1103/PhysRevLett.101.110501}.

\bibitem[Swingle(2012)]{Swingle:2009bg}
Brian Swingle.
\newblock {Entanglement Renormalization and Holography}.
\newblock \emph{Phys. Rev. D}, 86:\penalty0 065007, 2012.
\newblock \doi{10.1103/PhysRevD.86.065007}.

\bibitem[Rovelli(1998)]{Rovelli:1997yv}
Carlo Rovelli.
\newblock {Loop quantum gravity}.
\newblock \emph{Living Rev. Rel.}, 1:\penalty0 1, 1998.
\newblock \doi{10.12942/lrr-1998-1}.

\bibitem[Ashtekar and Lewandowski(2004)]{Ashtekar:2004eh}
Abhay Ashtekar and Jerzy Lewandowski.
\newblock {Background independent quantum gravity: A Status report}.
\newblock \emph{Class. Quant. Grav.}, 21:\penalty0 R53, 2004.
\newblock \doi{10.1088/0264-9381/21/15/R01}.

\bibitem[Li et~al.(2019)Li, Li, Han, Lu, Zhou, Ruan, Long, Wan, Lu, Zeng,
  et~al.]{li2019quantum}
Keren Li, Youning Li, Muxin Han, Sirui Lu, Jie Zhou, Dong Ruan, Guilu Long,
  Yidun Wan, Dawei Lu, Bei Zeng, et~al.
\newblock Quantum spacetime on a quantum simulator.
\newblock \emph{Communications Physics}, 2\penalty0 (1):\penalty0 122, 2019.

\bibitem[Mielczarek(2019)]{Mielczarek:2018jsh}
Jakub Mielczarek.
\newblock {Spin Foam Vertex Amplitudes on Quantum Computer - Preliminary
  Results}.
\newblock \emph{Universe}, 5\penalty0 (8):\penalty0 179, 2019.
\newblock \doi{10.3390/universe5080179}.

\bibitem[Zhang et~al.(2020)Zhang, Huang, Song, Guo, Song, Dong, Wang, Hekang,
  Han, Wang, et~al.]{zhang2020observation}
Pengfei Zhang, Zichang Huang, Chao Song, Qiujiang Guo, Zixuan Song, Hang Dong,
  Zhen Wang, Li~Hekang, Muxin Han, Haohua Wang, et~al.
\newblock Observation of two-vertex four-dimensional spin foam amplitudes with
  a 10-qubit superconducting quantum processor.
\newblock \emph{arXiv preprint arXiv:2007.13682}, 2020.

\bibitem[Mielczarek(2021)]{Mielczarek:2021xik}
Jakub Mielczarek.
\newblock {Prelude to Simulations of Loop Quantum Gravity on Adiabatic Quantum
  Computers}.
\newblock \emph{Front. Astron. Space Sci.}, 8:\penalty0 95, 2021.
\newblock \doi{10.3389/fspas.2021.571282}.

\bibitem[Czelusta and Mielczarek(2021)]{Czelusta:2020ryq}
Grzegorz Czelusta and Jakub Mielczarek.
\newblock {Quantum simulations of a qubit of space}.
\newblock \emph{Phys. Rev. D}, 103\penalty0 (4):\penalty0 046001, 2021.
\newblock \doi{10.1103/PhysRevD.103.046001}.

\bibitem[Feller and Livine(2016)]{Feller:2015yta}
Alexandre Feller and Etera~R. Livine.
\newblock {Ising Spin Network States for Loop Quantum Gravity: a Toy Model for
  Phase Transitions}.
\newblock \emph{Class. Quant. Grav.}, 33\penalty0 (6):\penalty0 065005, 2016.
\newblock \doi{10.1088/0264-9381/33/6/065005}.

\bibitem[Czelusta and Mielczarek(2023)]{czelusta2023quantum}
Grzegorz Czelusta and Jakub Mielczarek.
\newblock Quantum circuits for the ising spin networks.
\newblock \emph{Physical Review D}, 108\penalty0 (8):\penalty0 086027, 2023.

\bibitem[Maldacena(1998)]{Maldacena:1997re}
Juan~Martin Maldacena.
\newblock {The Large N limit of superconformal field theories and
  supergravity}.
\newblock \emph{Adv. Theor. Math. Phys.}, 2:\penalty0 231--252, 1998.
\newblock \doi{10.4310/ATMP.1998.v2.n2.a1}.

\bibitem[Han and Hung(2017)]{PhysRevD.95.024011}
Muxin Han and Ling-Yan Hung.
\newblock Loop quantum gravity, exact holographic mapping, and holographic
  entanglement entropy.
\newblock \emph{Phys. Rev. D}, 95:\penalty0 024011, Jan 2017.
\newblock \doi{10.1103/PhysRevD.95.024011}.
\newblock URL \url{https://link.aps.org/doi/10.1103/PhysRevD.95.024011}.

\bibitem[Cirac et~al.(2021)Cirac, Perez-Garcia, Schuch, and
  Verstraete]{Cirac:2020obd}
J.~Ignacio Cirac, David Perez-Garcia, Norbert Schuch, and Frank Verstraete.
\newblock {Matrix product states and projected entangled pair states: Concepts,
  symmetries, theorems}.
\newblock \emph{Rev. Mod. Phys.}, 93\penalty0 (4):\penalty0 045003, 2021.
\newblock \doi{10.1103/RevModPhys.93.045003}.

\bibitem[East et~al.(2023)East, Alonso-Linaje, and Park]{east2023all}
Richard~DP East, Guillermo Alonso-Linaje, and Chae-Yeun Park.
\newblock All you need is spin: Su (2) equivariant variational quantum circuits
  based on spin networks.
\newblock \emph{arXiv preprint arXiv:2309.07250}, 2023.

\bibitem[Mielczarek and Trze\'sniewski(2020)]{Mielczarek:2019srn}
Jakub Mielczarek and Tomasz Trze\'sniewski.
\newblock {Gauge fields and quantum entanglement}.
\newblock \emph{Phys. Lett. B}, 810:\penalty0 135808, 2020.
\newblock \doi{10.1016/j.physletb.2020.135808}.

\bibitem[Choi(1975)]{CHOI1975285}
Man-Duen Choi.
\newblock Completely positive linear maps on complex matrices.
\newblock \emph{Linear Algebra and its Applications}, 10\penalty0 (3):\penalty0
  285--290, 1975.
\newblock ISSN 0024-3795.
\newblock \doi{https://doi.org/10.1016/0024-3795(75)90075-0}.
\newblock URL
  \url{https://www.sciencedirect.com/science/article/pii/0024379575900750}.

\bibitem[Jamio{\l}kowski(1972)]{JAMIOLKOWSKI1972275}
A.~Jamio{\l}kowski.
\newblock Linear transformations which preserve trace and positive
  semidefiniteness of operators.
\newblock \emph{Reports on Mathematical Physics}, 3\penalty0 (4):\penalty0
  275--278, 1972.
\newblock ISSN 0034-4877.
\newblock \doi{https://doi.org/10.1016/0034-4877(72)90011-0}.
\newblock URL
  \url{https://www.sciencedirect.com/science/article/pii/0034487772900110}.

\bibitem[Bianchi and Livine(2023)]{bianchi2023loop}
Eugenio Bianchi and Etera~R Livine.
\newblock Loop quantum gravity and quantum information.
\newblock \emph{arXiv preprint arXiv:2302.05922}, 2023.

\bibitem[Khatri et~al.(2019)Khatri, LaRose, Poremba, Cincio, Sornborger, and
  Coles]{khatri2019quantum}
Sumeet Khatri, Ryan LaRose, Alexander Poremba, Lukasz Cincio, Andrew~T
  Sornborger, and Patrick~J Coles.
\newblock Quantum-assisted quantum compiling.
\newblock \emph{Quantum}, 3:\penalty0 140, 2019.

\bibitem[Knuth(1997)]{knuth1997art}
Donald~Ervin Knuth.
\newblock \emph{The art of computer programming}, volume~3.
\newblock Pearson Education, 1997.

\bibitem[Chen and Livine(2021)]{chen2021loop}
Qian Chen and Etera~R Livine.
\newblock Loop quantum gravity’s boundary maps.
\newblock \emph{Classical and Quantum Gravity}, 38\penalty0 (15):\penalty0
  155019, 2021.

\bibitem[Gross et~al.(2007)Gross, Audenaert, and Eisert]{gross2007evenly}
David Gross, Koenraad Audenaert, and Jens Eisert.
\newblock Evenly distributed unitaries: On the structure of unitary designs.
\newblock \emph{Journal of mathematical physics}, 48\penalty0 (5), 2007.

\bibitem[Dankert et~al.(2009)Dankert, Cleve, Emerson, and
  Livine]{dankert2009exact}
Christoph Dankert, Richard Cleve, Joseph Emerson, and Etera Livine.
\newblock Exact and approximate unitary 2-designs and their application to
  fidelity estimation.
\newblock \emph{Physical Review A—Atomic, Molecular, and Optical Physics},
  80\penalty0 (1):\penalty0 012304, 2009.

\bibitem[Colafranceschi et~al.(2022)Colafranceschi, Chirco, and
  Oriti]{colafranceschi2022holographic}
Eugenia Colafranceschi, Goffredo Chirco, and Daniele Oriti.
\newblock Holographic maps from quantum gravity states as tensor networks.
\newblock \emph{Physical Review D}, 105\penalty0 (6):\penalty0 066005, 2022.

\bibitem[Duda(2023)]{duda2023two}
Jarek Duda.
\newblock Two-way quantum computers adding cpt analog of state preparation.
\newblock \emph{arXiv preprint arXiv:2308.13522}, 2023.

\bibitem[Czelusta et~al.(2024)Czelusta, Verma, and
  Wanjalkar]{czelusta2024grover}
Grzegorz Czelusta, Dev~Rishi Verma, and Govind Wanjalkar.
\newblock Grover's algorithm on two-way quantum computer.
\newblock \emph{arXiv preprint arXiv:2406.09450}, 2024.

\bibitem[Cerezo et~al.(2021)Cerezo, Sone, Volkoff, Cincio, and
  Coles]{cerezo2021cost}
Marco Cerezo, Akira Sone, Tyler Volkoff, Lukasz Cincio, and Patrick~J Coles.
\newblock Cost function dependent barren plateaus in shallow parametrized
  quantum circuits.
\newblock \emph{Nature communications}, 12\penalty0 (1):\penalty0 1791, 2021.

\bibitem[Kingma(2014)]{kingma2014adam}
Diederik~P Kingma.
\newblock Adam: A method for stochastic optimization.
\newblock \emph{arXiv preprint arXiv:1412.6980}, 2014.

\bibitem[Mitarai et~al.(2018)Mitarai, Negoro, Kitagawa, and
  Fujii]{mitarai2018quantum}
Kosuke Mitarai, Makoto Negoro, Masahiro Kitagawa, and Keisuke Fujii.
\newblock Quantum circuit learning.
\newblock \emph{Physical Review A}, 98\penalty0 (3):\penalty0 032309, 2018.

\bibitem[Schuld et~al.(2019)Schuld, Bergholm, Gogolin, Izaac, and
  Killoran]{schuld2019evaluating}
Maria Schuld, Ville Bergholm, Christian Gogolin, Josh Izaac, and Nathan
  Killoran.
\newblock Evaluating analytic gradients on quantum hardware.
\newblock \emph{Physical Review A}, 99\penalty0 (3):\penalty0 032331, 2019.

\bibitem[Czelusta(2024)]{czelusta2024github}
G.~Czelusta.
\newblock Tensor network representation of ising spin networks, 2024.
\newblock URL
  \url{https://github.com/Quantum-Cosmos-Lab/Tensor-network-representation-of-Ising-spin-networks}.

\bibitem[Zyczkowski and Sommers(2001)]{zyczkowski2001induced}
Karol Zyczkowski and Hans-J{\"u}rgen Sommers.
\newblock Induced measures in the space of mixed quantum states.
\newblock \emph{Journal of Physics A: Mathematical and General}, 34\penalty0
  (35):\penalty0 7111, 2001.

\bibitem[Bergholm et~al.(2018)Bergholm, Izaac, Schuld, Gogolin, Ahmed, Ajith,
  Alam, Alonso-Linaje, AkashNarayanan, Asadi, et~al.]{bergholm2018pennylane}
Ville Bergholm, Josh Izaac, Maria Schuld, Christian Gogolin, Shahnawaz Ahmed,
  Vishnu Ajith, M~Sohaib Alam, Guillermo Alonso-Linaje, B~AkashNarayanan, Ali
  Asadi, et~al.
\newblock Pennylane: Automatic differentiation of hybrid quantum-classical
  computations.
\newblock \emph{arXiv preprint arXiv:1811.04968}, 2018.

\bibitem[Livine(2018)]{livine2018intertwiner}
Etera~R Livine.
\newblock Intertwiner entanglement on spin networks.
\newblock \emph{Physical Review D}, 97\penalty0 (2):\penalty0 026009, 2018.

\bibitem[Gozzini and Vidotto(2021)]{Gozzini:2019nbo}
Francesco Gozzini and Francesca Vidotto.
\newblock {Primordial Fluctuations From Quantum Gravity}.
\newblock \emph{Front. Astron. Astrophys. Cosmol.}, 7:\penalty0 629466, 2021.
\newblock \doi{10.3389/fspas.2020.629466}.

\bibitem[Czelusta and Mielczarek(2022)]{Czelusta:2021jro}
Grzegorz Czelusta and Jakub Mielczarek.
\newblock {Quantum variational solving of the Wheeler-DeWitt equation}.
\newblock \emph{Phys. Rev. D}, 105\penalty0 (12):\penalty0 126005, 2022.
\newblock \doi{10.1103/PhysRevD.105.126005}.

\bibitem[Grabowska et~al.(2024)Grabowska, Kane, and Bauer]{grabowska2024fully}
Dorota~M Grabowska, Christopher~F Kane, and Christian~W Bauer.
\newblock A fully gauge-fixed su (2) hamiltonian for quantum simulations.
\newblock \emph{arXiv preprint arXiv:2409.10610}, 2024.

\bibitem[Cepollaro et~al.(2024)Cepollaro, Chirco, Cuffaro, Esposito, and
  Hamma]{Cepollaro:2024qln}
Simone Cepollaro, Goffredo Chirco, Gianluca Cuffaro, Gianluca Esposito, and
  Alioscia Hamma.
\newblock {Stabilizer entropy of quantum tetrahedra}.
\newblock \emph{Phys. Rev. D}, 109\penalty0 (12):\penalty0 126008, 2024.
\newblock \doi{10.1103/PhysRevD.109.126008}.

\end{thebibliography}

\end{document}